\documentclass[12pt]{iopart}
\usepackage[latin1]{inputenc}
\usepackage{amssymb}
\usepackage[pdftex]{graphicx}
\usepackage{float}
\usepackage{overpic}
\usepackage{soul}
\usepackage{color}

\providecommand{\U}[1]{\protect\rule{.1in}{.1in}}
\newcommand{\lr}{\scriptscriptstyle \! \nearrow \!\!\!\!\!\! \nwarrow \;}
\newcommand{\eqref}[1]{(\ref{#1})}

\newcommand{\vecr}{\mathbf{r}} 
\newcommand{\ie}{\textit{i.e. }}
\newcommand{\eg}{\textit{e.g. }}
\renewcommand{\i}{\mathrm{i}}

\begin{document}
\title[General Boundary Conditions for Quasiclassical Theory in the Diffusive Limit]{General Boundary Conditions for Quasiclassical Theory of Superconductivity in the Diffusive Limit: Application to Strongly Spin-polarized Systems\footnote{
This is an author-created, un-copyedited version of an article published in 
New Journal of Physics: 
{\it New J. Phys.} {\bf 17} (2015) 083037 (1-21). IOP Publishing Ltd is not responsible for any errors or omissions in this version of the manuscript or any version derived from it. The Version of Record is available online at doi:10.1088/1367-2630/17/8/083037.\\
Corresponding author: M. Eschrig: matthias.eschrig@rhul.ac.uk}}
\author{M. Eschrig$^{1,2}$, A. Cottet$^{3}$, W. Belzig$^4$, J. Linder$^2$}
\address{$^1$ Department of Physics, Royal Holloway, University of London, Egham, Surrey TW20 0EX, United Kingdom}
\address{$^2$ Department of Physics, Norwegian University of Science and Technology, N-7491 Trondheim, Norway}
\address{$^3$ 
Laboratoire Pierre Aigrain, Ecole Normale Sup\'erieure-PSL Research University, CNRS, Universit\'e Pierre et Marie Curie-Sorbonne Universit\'es, Universit\'e Paris Diderot-Sorbonne Paris Cit\'e, 24 rue Lhomond, F-75231 Paris Cedex 05, France }
\address{$^4$ Department of Physics, University of Konstanz, D-78457 Konstanz, Germany}

\date{22.04.2015}

\begin{abstract}
Boundary conditions in quasiclassical theory of superconductivity are of crucial importance for describing proximity effects in heterostructures between different materials. Although they have been derived for the ballistic case in full generality, corresponding boundary conditions for the diffusive limit, described by Usadel theory, have been lacking for interfaces involving strongly spin-polarized materials, such as \eg half-metallic ferromagnets. 
Given the current intense research in the emerging field of superconducting spintronics, the formulation of appropriate boundary conditions for the Usadel theory of diffusive superconductors in contact with strongly spin-polarized ferromagnets for arbitrary transmission probability and arbitrary spin-dependent interface scattering phases has been a burning open question.  Here we close this gap and 
derive the full boundary conditions for quasiclassical Green functions in the diffusive limit, valid for any value of spin polarization, transmission probability, and spin-mixing angles (spin-dependent scattering phase shifts). It allows also for complex spin textures across the interface and for channel off-diagonal scattering (a necessary ingredient when the numbers of channels on the two sides of the interface differ). As an example we derive expressions for the proximity effect in diffusive systems involving half-metallic ferromagnets. 
In a superconductor/half-metal/superconductor Josephson junction we find $\phi_0$-junction behavior under certain interface conditions.
\end{abstract}
\maketitle 

\section{Introduction}
\let\ootimes\otimes
\renewcommand{\otimes}{\circ}
 
Hybrid structures containing superconducting (S) and ferromagnetic (F) materials became a focus of nanoelectronic research because of their relevance for
spintronics applications as well as their potential impact on fundamental research 
\cite{Eschrig11,Eschrig15,Linder15}. 
Examples of successful developments include the 
discoveries of the $\pi$-junction \cite{Bulaevskii77,Buzdin82} in S/F/S Josephson devices \cite{Ryazanov01,Kontos02}, 
of odd-frequency superconductivity \cite{Berezinskii74} in S/F heterostructures \cite{Bergeret01,Kadigrobov01}, and of the indirect Josephson effect in S/half-metal/S junctions \cite{Eschrig03,eschrig_nphys_08}. Other recent topics of interest include the study of Majorana fermions at interfaces between superconductors and topological insulators \cite{Tanaka12} and at edges in superfluid $\,^3$He \cite{Volovik02,Roy08}, 
and the appearance of pure spin supercurrents in topological superconductors \cite{Vorontsov08}, and in S/FI-F-FI devices as a result of geometric phases \cite{grein_prl_09}.

The central subject in many of these studies is 
to understand how in the case of a superconductor
coupled to a ferromagnetic material superconducting
correlations penetrate into the ferromagnet, and how magnetic correlations penetrate into the superconductor \cite{Izyumov02,Golubov04,Eschrig04,Buzdin05,Bergeret05,Pokrovsky07}. 
A powerful method
to treat such problems is the quasiclassical theory of superconductivity
developed by Larkin and Ovchinnikov and by Eilenberger \cite{Eilenberger,Larkin}.
Within this theory \cite{Serene83,Rammer86,Belzig99,Eschrig01,Kopnin09} the quasiparticle motion is treated on 
a classical level, whereas the particle-hole  and the spin degrees
of freedom are treated quantum mechanically.
The transport equation, which is a first order matrix differential equation for the quasiclassical propagator, must be supplemented by physical boundary conditions in order to obtain a unique solution.

Whereas for the full microscopic Green functions, the Gor'kov Green functions \cite{Gorkov58}, such boundary conditions can be readily formulated (\eg in terms of interface scattering matrices or in terms of transfer matrices), this is a considerably more difficult task for quasiclassical Green functions. In quasiclassical theory only the information about the envelope functions of Bloch waves is retained, information about the phases of the waves is however missing.
Such envelope amplitudes can show jumps at interfaces, and one complex task is to calculate these jumps without knowing the full microscopic Green functions near the interface. 
Correspondingly, there is a long history of deriving boundary conditions for quasiclassical propagators, both for the Eilenberger equations, and their diffusive limit, the Usadel equations \cite{Usadel}.

For ballistic transport, described by the Eilenberger equations, such boundary conditions were first formulated for spin-inactive interfaces in pioneering work by Shelankov and by Zaitsev \cite{Shelankov84,Zaitsev84}, who showed 
the non-trivial fact that these jumps can be calculated using only the envelope functions. More general formulations were proposed subsequently \cite{Ashauer86,Zhang87,Nagai88,Millis88}, including a formulation in terms of interface scattering matrices by Millis, Rainer, and Sauls \cite{Millis88}.
All these formulations were implicit in terms of non-linear matrix equations, and problems arose in numerical implementations due to spurious (unphysical) additional solutions which must be eliminated. 
Progress was made with the help of Shelankov's projector formalism \cite{Shelankov80}, allowing for explicit formulations of boundary conditions in both equilibrium \cite{Yip97,Eschrig00,Shelankov00} and non-equilibrium \cite{Eschrig00} situations.
Further generalizations included spin-active interfaces, formulated for equilibrium \cite{Fogelstrom00} and for non-equilibrium \cite{Zhao04}, and interfaces with diffusive scattering characteristics \cite{Lueck03}. An alternative formulation in terms of quantum mechanical $t$-matrices \cite{Cuevas96} proved also fruitful \cite{Cuevas01,Huertas02,Eschrig03,Eschrig04,Kopu04,Graser07}. The latest formulation, in terms of interface scattering matrices, is able to include non-equilibrium phenomena, interfaces and materials with weak or strong spin polarization, multi-band systems, as well as disordered systems \cite{Eschrig09}.

For the diffusive limit a set of second order matrix differential equations has been derived by Usadel \cite{Usadel}. 
In contrast to the ballistic case, where boundary conditions have been formulated for a wide set of applications, boundary conditions for the diffusive limit have been formulated so far only in certain limiting cases.
The first formulation is by Kupriyanov and Lukichev, appropriate for the tunneling limit \cite{Kupriyanov88}. This was generalized to arbitrary transmission by Nazarov \cite{naz}. A major advance was done by Cottet \etal in formulating boundary conditions for Usadel equations appropriate for spin-polarized interfaces \cite{cottet}. These boundary conditions are valid in the limit of small transmission, spin polarization, and spin-dependent scattering phase shifts (this term is often used interchangeably with ``spin-mixing angles'' \cite{Tokuyasu88}). 
Subsequent formulations allowed for arbitrary spin polarization, although being restricted to small transmission and spin-dependent scattering \cite{Machon1,Machon2,Bergeret12}. In Ref. \cite{Bergeret12} the authors present ``heuristically'' deduced boundary conditions, which coincide with the ones used in Refs. \cite{Machon1,Machon2}.

Here we not only present the full derivation of the specific boundary conditions used in Refs. \cite{Machon1,Machon2,Bergeret12}, but go further and give a full solution of the problem.
With this, the long-standing problem of how to generalize Nazarov's formula for arbitrary transmission probability \cite{naz} to the case of spin-polarized systems with arbitrary spin polarization and arbitrary spin dependent scattering phases is solved. Our boundary conditions are general enough to allow for non-equilibrium situations within Keldysh formalism, as well as for complex interface spin textures. We reproduce as limiting cases all previously known formulations.

\section{Transport Equations}
The central quantity in quasiclassical theory of superconductivity
 \cite{Eilenberger,Larkin} is 
the quasiclassical Green function (``propagator'')
$\check{g}({\bf p}_F,{\bf R},E,t)$.
It describes quasiparticles with
energy $E$ (measured from the Fermi level) and momentum ${\bf p}_F$ 
moving along classical trajectories with direction given by the Fermi velocity
${\bf v}_F({\bf p}_F)$ in external potentials and self-consistent fields that are modulated by the slow spatial (${\bf R}$) and time ($t$) coordinates \cite{Serene83,Rammer86,Belzig99}.
The quasiclassical Green function is a functional of 
self-energies $\check\Sigma({\bf p}_F,{\bf R},E,t)$, which in general
include molecular fields, the superconducting order parameter 
$\Delta ({\bf p}_F,{\bf R},t)$,
impurity scattering, and the external potentials.
The quantum mechanical degrees of freedom of the quasiparticles show up in the
matrix structure of the quasiclassical propagator and the self-energies.
It is convenient to formulate the theory using
2$\times$2 matrices in Keldysh space \cite{Keldysh} (denoted by a ``check'' accent), the elements of
which in turn are  2$\times$2 Nambu-Gor'kov matrices \cite{Gorkov58,Nambu} in 
particle-hole (denoted by a ``hat'' accent) space.
The structure of the
propagators and self-energies in Keldysh-space is 
\numparts
\begin{eqnarray}
\check g= \left(
\begin{array}{cc}
\hat g^R & \hat g^K \\
0 & \hat g^A
\end{array}
\right)_{\!\rm kel},
\quad 
\label{sigma}
\check{\Sigma}=\left( \begin{array}{cc} \hat{\Sigma}^R & \hat{\Sigma}^K \\
0 & \hat{\Sigma}^A \end{array} \right)_{\! \rm kel},
\end{eqnarray}
where the superscripts $R$, $A$, and $K$ refer to retarded, advanced, and Keldysh components, respectively, and
with the particle-hole space structure
\footnote{
For the definitions of all Green functions in this paper we use a basis of fermion field operators in Nambu $\ootimes$ spin-space as
$\Psi(\vecr,t) = [\psi_\uparrow(\vecr,t),  \psi_\downarrow(\vecr,t),  \psi_\uparrow(\vecr,t)^\dag,  \psi_\downarrow(\vecr,t)^\dag]^T$ .
}
\begin{eqnarray}
\label{gl_green3}
    \hat{g}^{R,A}=\!
    \left( \begin{array}{cc} g^{R,A} & f^{R,A} \\ \tilde{f}^{R,A} & \tilde {g}^{R,A}
      \end{array} \right)_{\! \rm ph},\quad \hat{g}^{K}=\!
    \left( \begin{array}{cc} \; \, g^K & \; \, f^K \\ -\tilde{f}^K & -\tilde {g}^K
      \end{array} \right)_{\! \rm ph}
\end{eqnarray}
for Green functions, and
\begin{eqnarray}
\label{gl_self3}
    \hat{\Sigma}^{R,A}=\!
    \left( \begin{array}{cc} \Sigma^{R,A} & \Delta^{R,A} \\ \tilde{\Delta}^{R,A} & 
     \tilde {\Sigma}^{R,A}
      \end{array} \right)_{\! \rm ph},\quad \hat{\Sigma}^{K}=\!
    \left( \begin{array}{cc} \;\,  \Sigma^K & \; \, \Delta^K \\ -\tilde{\Delta}^K & 
      -\tilde {\Sigma}^K
      \end{array} \right)_{\! \rm ph}
\end{eqnarray}
\endnumparts
for self-energies.
For spin-degenerate trajectories (\ie in systems with weak or no spin-polarization)
the elements of the 2$\times$2 Nambu-Gor'kov matrices
are 2$\times$2 matrices in spin space, \eg $g^R=g^R_{ab}$ with $a,b\in \{\uparrow,
\downarrow\}$, and similarly for others. In strongly spin-polarized ferromagnets the elements of the 2$\times$2 Nambu-Gor'kov matrices are spin-scalar 
(due to very fast spin-dephasing in a strong exchange field),
and the system must be described within the preferred quantization direction given by the internal exchange field.
The terms ``weak'' and ``strong'' refer to the spin-splitting of the energy bands being comparable to the superconducting gap or to the band width, respectively.
In writing Eqs. \eqref{sigma}-\eqref{gl_self3} we used general symmetries, which are accounted for
by the ``tilde'' operation,
\begin{equation} 
\label{tilde}
    \tilde{X}({\bf p}_F,{\bf R},E,t)=X(-{\bf p}_F,{\bf R},-E,t)^\ast.
\end{equation}
Retarded (advanced) functions can be analytically continued into the upper (lower) complex energy half plane, in which case the relation is modified to $\tilde{X}({\bf p}_F,{\bf R},E,t)=X(-{\bf p}_F,{\bf R},-E^\ast,t)^\ast$ with complex $E$.

The quasiclassical Green functions satisfy the Eilenberger-Larkin-Ovchin\-nikov 
transport equation and normalization condition
\begin{equation}
\left[E \check \tau_3 - \check \Sigma ,
\check g
\right]_{\otimes} +
\i \hbar {\bf v}_F \cdot \nabla
\check g=\check 0,
\quad
\check g \otimes \check g = -\pi^2 \check 1.
\label{eilen}
\end{equation}
The non-commutative product $\otimes$ combines
matrix multiplication with a convolution over the internal energy-time variables in Wigner coordinate representation,
\begin{equation}
(\check A \otimes \check B)(E,t) \equiv
e^{\frac{\i}{2} (\partial_E^A\partial_t^B-\partial_t^A\partial_E^B)} \check A(E,t) \check B(E,t),
\end{equation}
and $\check \tau_3=\hat \tau_3 \check 1$, where $\hat \tau_3$ is a Pauli matrix in
particle-hole space. Here and below, $\left[A,B \right]_\otimes \equiv 
A\otimes B-B\otimes A$. The operation $\nabla$ acts on the variable ${\bf R}$.

The functional dependence of the quasiclassical propagator on the self-energies
is given in the form of self-consistency conditions.
For instance, for a weak-coupling, $s$-wave order parameter the condition reads
\begin{equation}
\hat \Delta ({\bf R},t) = 
V_{s}\int^{E_c}_{-E_c} \frac{dE }{4\pi \i}
\langle N_{F}({\bf p}_F)\hat f^{K}_s({\bf p}_F,{\bf R},E,t) \rangle_{{\bf p}_F},
\end{equation}
where $V_{s}$ is the $s$-wave part of the singlet pairing interaction, $N_F$ is the density of states per spin at the Fermi level, $\hat f^{K}_s$ is spin-singlet part of the the Keldysh component $\hat f^{K}$,
and $\langle
\hspace{2mm} \rangle_{{\bf p}_F}$ denotes averaging over the Fermi surface.
The cut-off energy $E_c$ is to be eliminated in favor of the superconducting transition
temperature in the usual manner.

When the quasiclassical Green function has been determined,
physical quantities of interest can be calculated.
For example, the current density at position ${\bf R}$ and time $t$ reads (with $e<0$ the electron charge)
\begin{equation}
{\bf j} ({\bf R},t) = e
\int^{\infty}_{-\infty} \frac{dE }{8\pi \i}
{\rm Tr} \langle N_{F}({\bf p}_F)
{\bf v}_F({\bf p}_F) \hat \tau_3 \hat g^{K}({\bf p}_F,{\bf R},E,t)\rangle_{{\bf p}_F}.
\label{densityofstates}
\end{equation}
The symbol Tr denotes a trace over the 2$\times$2 particle-hole space as well as
over 2$\times$2 spin space in the case of spin-degenerate trajectories.

In the dirty (diffusive) limit, strong scattering by non-magnetic impurities effectively averages the quasiclassical propagator over momentum directions.
The Green function may then be expanded in the small parameter $k_{\rm B}T_{c}\tau/\hbar$  ($\tau$ is the momentum relaxation time) following the standard procedure \cite{Usadel,Alexander85}
\begin{eqnarray}
\check{g}({\bf p}_F,{\bf R},E,t) \approx \check{G}({\bf R},E,t) + \check{g}^{(1)} ({\bf p}_F,{\bf R},E,t) \label{exp}
\end{eqnarray}
where the magnitude of $\check{g}^{(1)}$ is small compared to that of $\check{G}$.
The impurity self-energy  is related to an (in general anisotropic) lifetime function $\tau({\bf p}_F',{\bf p}_F)$ \cite{Alexander85}.
Substituting \eqref{exp} into \eqref{eilen}, 
multiplying with $N_F({\bf p}_F')\mbox{v}_{F,j}({\bf p}_F')\tau({\bf p}_F',{\bf p}_F)$,
averaging over  momentum directions, considering that $\check{\Sigma}' \tau /\hbar $ is small, where $\check{\Sigma}' $ is the self-energy reduced by the contribution due to non-magnetic impurity scattering,
 and using $\check G\otimes \check G=-\pi^2 \check 1$ and $\check G\otimes \check{g}^{(1)} + \check{g}^{(1)} \otimes \check G = \check 0$,
one obtains (we suppress here the arguments ${\bf R},E,t$)
\begin{equation}
\label{relation}
\left\langle N_F({\bf p}_F) \mbox{v}_{F,j}({\bf p}_F) \check{g}^{(1)}({\bf p}_F)\right\rangle_{{\bf p}_F} = N_F\sum_k
\frac{D_{jk}}{\i\pi} \check{G} \otimes \nabla_k \check{G}, 
\end{equation}
where $N_F=\langle N_F({\bf p}_F )\rangle_{{\bf p}_F}$
is the local density of states per spin at the Fermi level,
$\nabla_k=\partial/\partial R_k$, the summation is over $k\in \left\{x,y,z\right\}$, and 
\begin{equation}
D_{jk}= \frac{1}{N_F^2}\Big\langle\Big\langle
N_F({\bf p}_F' ) \mbox{v}_{F,j}({\bf p}_F') \, \tau ({\bf p}_F',{\bf p}_F)\, 
\mbox{v}_{F,k}({\bf p}_F) N_F({\bf p}_F) 
\Big\rangle_{{\bf p}_F}\Big\rangle_{{\bf p}_F'}
\end{equation}
is the diffusion constant tensor.  For isotropic systems, $D_{jk}=D\delta_{jk}$.
The Usadel Green function $\check{G}$ obeys the following 
transport equation and normalization condition \cite{Usadel},
\begin{eqnarray} 
\label{gl_usdl}
    \left[ E \hat{\tau}_3 \check{1} -\check{\Sigma}_0\, , \, 
    \check{G} \right]_\otimes &+&\sum_{jk}\frac{\hbar D_{jk}}{\pi} \nabla_j \left( \check{G} 
	\otimes \nabla_k \check{G}
    \right) = \check 0, \quad
\check G \otimes \check G = -\pi^2 \check 1, 
\end{eqnarray} 
where $\check \Sigma_0=\langle N_F({\bf p}_F) \check \Sigma'({\bf p}_F)\rangle_{{\bf p}_F}/N_F$.
The Usadel propagator $\check G$ is a functional of $\check \Sigma_0$.

The structures of $\check G$ and $\check \Sigma_0$ are the same as in
Eqs. \eqref{sigma}-\eqref{gl_self3} (with $\check G$ replacing $\check g$ and $\Sigma_0$ replacing $\Sigma $). Eq. \eqref{tilde} is replaced by
\begin{equation} 
\label{ustilde}
    \tilde{X}({\bf R},E,t)=X({\bf R},-E,t)^\ast.
\end{equation}
The current density for diffusive systems is obtained from Eqs. \eqref{relation} and \eqref{densityofstates}, and is given by
\begin{equation}
j_i({\bf R},t) = -e \sum_k 
\int^{\infty}_{-\infty} \frac{dE }{8\pi^2}
{\rm Tr} N_F D_{ik} \hat \tau_3 [\check G ({\bf R},E,t)\otimes \nabla_k \check G({\bf R},E,t)]^{K} .
\label{densityofstatesdiff}
\end{equation}
A vector potential ${\bf A}({\bf R},t)$ enters in a gauge invariant manner by replacing the spatial derivative operators in all expressions by (see \eg  \cite{Alexander85,Tanaka09})
\begin{equation}
\nabla_{i} \hat X \to \hat \partial_i \otimes \hat X \equiv \nabla_{i} \hat X -\i \left[ \frac{e}{\hbar }\hat \tau_3 A_i,\hat X\right]_{\otimes}.
\end{equation}

Finally, the case of a strongly spin-polarized itinerant ferromagnet with superconducting correlations (\eg due to the proximity effect when in contact with a superconductor) can be treated by quasiclassical theory as well \cite{Eschrig03,Eschrig04,Kopu04}. In this case, when the spin-splitting of the energy bands
is comparable to the band width of the two spin bands, there exist two well separated fully spin-polarized Fermi surfaces in the system, and the length scale associated with $\hbar/|{\bf p}_{F\uparrow}-{\bf p}_{F\downarrow} |$ is much shorter than the coherence length scale in the ferromagnet.
Equal-spin correlations stay still coherent over long distance in such a system;
$\uparrow\downarrow $ and $\downarrow\uparrow$ correlations are, however, incoherent and thus negligible within quasiclassical approximation. 
Fermi velocity, density of states,
diffusion constant tensor, and coherence length all become spin-dependent.
The quasiclassical propagator is then spin-scalar for each trajectory, with either all elements $\uparrow\uparrow $ or all elements $\downarrow\downarrow $ depending on the spin Fermi surface the trajectory corresponds to.
Eilenberger equation and Usadel equation have the same form as before for each separate spin band.
The spin-resolved current densities are given in the ballistic case by
\begin{equation}
{\bf j}_{\uparrow}=e 
\int^{\infty}_{-\infty} \frac{dE }{8\pi \i} {\rm Tr} 
\big\langle N_{F\uparrow} {\bf v}_{F\uparrow} \hat \tau_3 \hat g^K_{\uparrow\uparrow} \big\rangle_{{\bf p}_{F\uparrow}} ,
\label{currentstrong}
\end{equation}
and in the diffusive case by
\begin{equation}
j_{k\uparrow}=
-e \sum_k \int^{\infty}_{-\infty} \frac{dE }{8\pi^2} {\rm Tr}
N_{F\uparrow} D_{\uparrow kj} \hat \tau_3 \left[ \check G_{\uparrow\uparrow}\otimes \nabla_j \check G_{\uparrow\uparrow} \right]^K,
\label{diffcurrentstrong}
\end{equation}
and analogously for spin down.

For heterostructures, the above equations must be supplemented with
boundary conditions at the interfaces. A practical formulation of boundary conditions for diffusive systems valid for arbitrary transmission and spin polarization is the goal of this paper.

\section{Boundary Conditions }
\subsection{Interface Scattering Matrix}
We formulate boundary conditions at an interface in terms of the normal-state interface scattering matrix $\hat {\bf S}$ \cite{Lambert91,Takane92,Beenakker92}, connecting incoming with outgoing Bloch waves on either side of the interface with each other. We use the notation
\begin{equation}
\label{SM0}
\hat {\bf S}=
\left(
\begin{array}{cc}
\hat {\bf S }_{11} & \quad \hat {\bf S}_{12} \\  \hat {\bf S}_{21} & -\hat {\bf S}_{22}
\end{array}
\right)_{\!\lr},
\end{equation}
where $1$ and $2$ refer to the two sides of the interface, and the subscript label $\; \lr \!\! $ indicates that the 2x2 matrix structure refers to reflection and transmission amplitudes at an interface. 
The components $\hat {\bf S}_{ij}$ 
are matrices in particle-hole space as well as in scattering channel space 
(\eg scattering channels for ballistic transport 
would be parameterized by the Fermi momenta of incoming and outgoing Bloch waves). 
Each element in 2$\times$2 particle hole space is in turn
a matrix in combined spin and channel space, i.e. the number of incoming directions (assumed to be equal to the number of outgoing directions due to particle conservation) gives the dimension in channel space. The dimension in spin space is for spin-degenerate channels 2 and for spin-scalar channels 1.

If time-reversal symmetry is preserved, Kramers degeneracy requires that each 
element of the scattering matrix has a 2x2 spin (or more general: pseudo-spin) structure (as it connects doubly degenerate scattering channels on either side of the interface). For spin-polarized interfaces (\eg ferromagnetic or with Rashba spin-orbit coupling) the scattering matrix is not spin-degenerate. However if the splitting of the spin-degeneracy is on the energy scale of the superconducting gap, it can be neglected within the precision of quasiclassical theory of superconductivity. On the other hand, if the lifting of the spin-degeneracy of energy bands is comparable to the Fermi energy, the degeneracy of the scattering channels must be lifted as well in order to achieve consistency within quasiclassical theory.
For definiteness, we denote the dependence on the scattering channels by indices $n,n'$:
\begin{equation}
[\hat {\bf S}_{\alpha \beta}]_{nn'},
\end{equation}
even for the ballistic case for which 
$[\hat {\bf S}_{\alpha \beta}]_{nn'} \equiv \hat {\bf S}_{\alpha \beta} ({\bf p}_{F,n},{\bf k}_{F,n'})$.

As shown in \ref{app1} and \ref{app2},
the scattering matrix for an interface can be written in polar decomposition in full generality as
\begin{equation}
\hat {\bf S}=
\left( \begin{array}{cc}
\sqrt{1-C C^\dagger }& C\\ C^\dagger &
-\sqrt{1-C^\dagger C}
\end{array} \right)_{\! \lr}
\left(
\begin{array}{cc} {\cal S} & 0\\ 0& \breve{\cal S}   \end{array}
\right)_{\! \lr}
\end{equation}
with unitary matrices ${\cal S}$ and $\breve{\cal S}$, and a transmission matrix $C$. All are matrices in particle-hole space, scattering channel space, and possibly (pseudo-)spin space. The above decomposition divides the scattering matrix into a Hermitian part and a unitary part. From this decomposition, we can define the auxiliary scattering matrix 
\begin{eqnarray}
\label{SV0}
\hat {\bf S}_0&=&
\left(
\begin{array}{cc} {\cal S} & 0\\ 0& \breve{\cal S} \end{array}
\right)_{\! \lr}
,
\end{eqnarray}
which retains all the phase information during reflection on both sides of the interface, and has zero transmission components.
The decomposition
is uniquely defined when there are no zero-reflection singular values (we will assume here that always a small non-zero reflection takes place for each transmission channel; perfectly transmitting channels can always be treated separately as the corresponding boundary conditions are trivial).
For the matrix $C$ we introduce the parameterization 
\begin{equation}
C=\left(1+tt^\dagger \right)^{-1} 2t ,
\label{C}
\end{equation}
(see \ref{app3})
which is uniquely defined when all singular values of $t$ are in the interval $[0,1]$ (which is required in order to ensure non-negative reflection singular values).
We define for notational simplification ``hopping amplitude'' matrices 
\begin{eqnarray}
\pi \tau_{12} = t\breve{\cal S},\quad
\pi \tau_{21}=t^\dagger {\cal S} , 
\label{tau}
\end{eqnarray}
as well as unitary matrices 
\begin{eqnarray}
S_1={\cal S},\qquad S_2=\breve{\cal S}. 
\end{eqnarray}
In terms of those, obviously the relation
\begin{eqnarray}
\label{tausymm}
\tau_{\alpha \bar \alpha } = S_\alpha (\tau_{\bar \alpha \alpha })^\dagger S_{\bar \alpha }
\end{eqnarray}
holds, where $(\alpha, \bar \alpha )\in \{(1,2),(2,1)\}$, and the labels 1 and 2 refer to the respective sides of the interface. 
Here, and below, the Hermitian conjugate operation involves a transposition in channel indices.
The particle-hole structures of the surface scattering matrix and the
hopping amplitude are given by,
\begin{eqnarray}
\hat S_{\alpha}&=&\left(
\begin{array}{cc}
S_{\alpha}& 0 \\
0 & (\tilde S_{\alpha})^\dagger
\end{array}
\right)_{\! \rm ph}, \qquad
\hat \tau_{\alpha\bar \alpha }=\left(
\begin{array}{cc}
\tau_{\alpha \bar \alpha } & 0 \\
0 & 
(\tilde \tau_{\bar \alpha \alpha })^\dagger
\end{array}
\right)_{\!\rm ph},
\end{eqnarray}
with 
\begin{eqnarray}
\, [\tilde S_{\alpha} ]_{nn'}&=& [S_{\alpha}]_{\bar n\bar n'}^{\ast}, \quad
\, [\tilde \tau_{\alpha\bar\alpha }]_{nn'}= [\tau_{\alpha \bar \alpha }]_{\bar n\bar n'}^{\ast},
\end{eqnarray}
where $\bar n$ and $\bar n'$ denote mutually conjugated channels, \eg defined by
${\bf p}_{F,\bar n'}\equiv -{\bf k}_{F,n'}$ and ${\bf k}_{F,\bar n} \equiv -{\bf p}_{F,n}$.
Finally, the Keldysh structure of these quantities is
\begin{eqnarray}
\check S_{\alpha}&=&
\left( \begin{array}{cc}
\hat S_{\alpha}^{R}& 0 \\
0 & (\hat S_{\alpha}^{A})^\dagger
\end{array}
\right)_{\! \rm kel}
\equiv
\left( \begin{array}{cc}
\hat S_{\alpha}& 0 \\
0 & \hat S_{\alpha}
\end{array}
\right)_{\! \rm kel}
, \\
\check \tau_{\alpha\bar \alpha }&=&\left(
\begin{array}{cc}
\hat \tau_{\alpha \bar \alpha }^{R} & 0 \\
0 & 
(\hat \tau_{\bar \alpha \alpha }^{A})^\dagger
\end{array} \right)_{\!\rm kel } 
\equiv
\left( \begin{array}{cc}
\hat \tau_{\alpha \bar \alpha } & 0 \\
0 & 
\hat \tau_{\alpha \bar \alpha }
\end{array} \right)_{\!\rm kel } 
\end{eqnarray}
(the additional Hermitian conjugate in these equations is due to the fact that advanced Green functions have the roles of ``incoming'' and ``outgoing'' momentum directions interchanged compared to retarded Green functions; this is similar to the additional Hermitian conjugate appearing for hole components in particle-hole space).
Thus, the Keldysh matrix structure for $\check S_{\alpha}$ and $\check \tau_{\alpha\bar \alpha }$ is trivial (proportional to unit matrix).
The full normal-state scattering matrix is diagonal in particle-hole and in Keldysh space, with reflection 
components
\begin{eqnarray}
\check {\bf S}_{\alpha\alpha}&=& 
(1+\pi^2\check \tau_{\alpha\bar \alpha } \check \tau_{\alpha \bar\alpha }^\dagger )^{-1} \; (1-\pi^2\check \tau_{\alpha\bar\alpha } \check \tau_{\alpha \bar\alpha }^\dagger)\;
\check S_\alpha ,
\end{eqnarray}
and with transmission components
\begin{eqnarray}
\check {\bf S}_{\alpha\bar\alpha }&=&
(1+\pi^2\check\tau_{\alpha\bar\alpha } \check \tau_{\alpha \bar\alpha }^\dagger)^{-1} \; 2\pi \check \tau_{\alpha\bar\alpha } .
\end{eqnarray}
Note that $\tau_{\alpha \bar \alpha } $ connects incoming with outgoing Bloch waves per definition (as the scattering matrix does).

\begin{figure}[t!]
\centering{
(a)\includegraphics[width=0.7\linewidth]{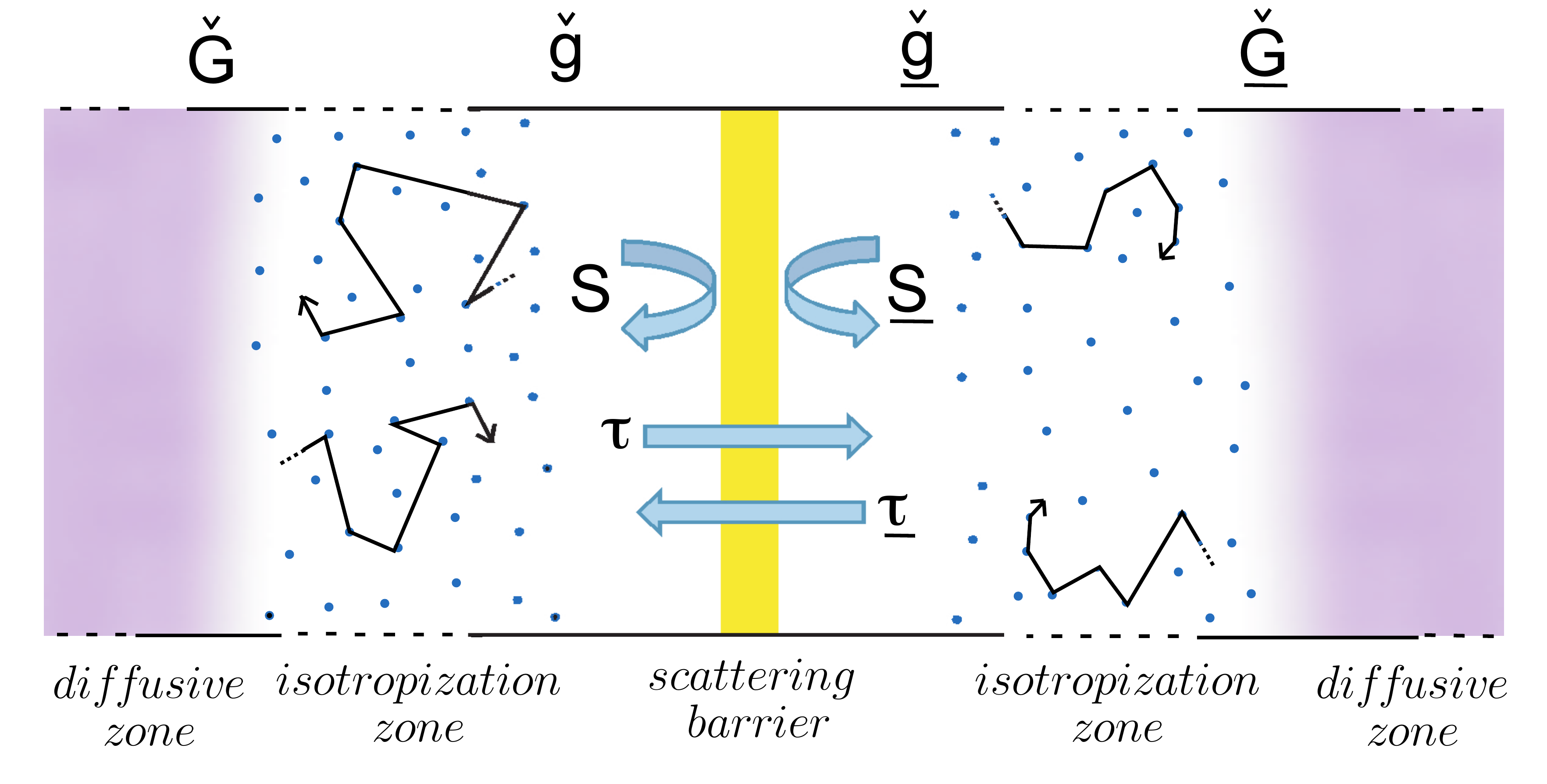}\\
\includegraphics[width=0.5\linewidth]{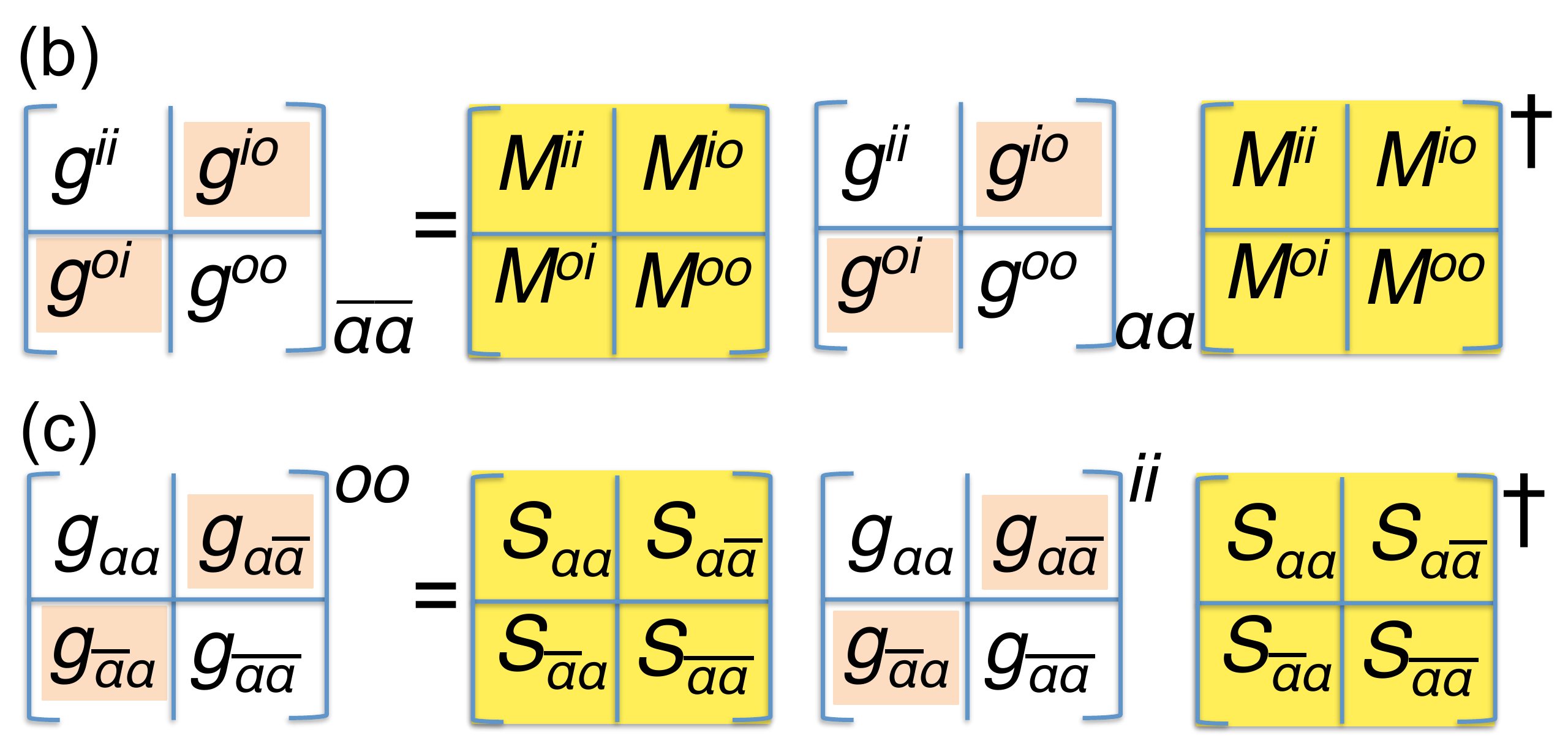} 
}
\caption{
(a):
Illustration of notation used in this paper.
(b) and (c): 
Structure of boundary condition with transfer matrices ${\bf M}$ in (b), and with scattering matrices ${\bf S}$ in (c) (yellow). 
``Drone'' amplitudes in the propagators (orange fields) connect in (b) incoming ($i$) and outgoing ($o$) momentum directions, and in (c) the two sides, $\alpha $ and $\overline\alpha $, of the interface.
To obtain quasiclassical boundary conditions, Drone amplitudes in (b) and (c) must be eliminated.
In this paper we use formulation (c).
To connect to the notation in the main text, 
$g^{ii}_{\alpha\alpha}\equiv g^i$,
$g^{ii}_{\bar\alpha\bar\alpha}\equiv \underline{g}^i$,
$g^{oo}_{\alpha\alpha}\equiv g^o$, and
$g^{oo}_{\bar\alpha\bar \alpha}\equiv \underline{g}^o$.
}
\label{fig:notation}
\end{figure}

We will formulate the theory such that all equations are valid on either side of the interface. This allows us to drop the indices $\alpha, \bar \alpha $ for simplicity of notation by randomly choosing one side of the interface, and denoting quantities on the other side of the interface by underline. In particular, we will use 
\begin{eqnarray}
&&\check S_\alpha \equiv \check S,
\quad \check S_{\bar \alpha} \equiv \underline {\check S}, \quad
 \check \tau_{\alpha \bar \alpha}\equiv \check \tau ,\quad
 \check \tau_{\bar \alpha \alpha}\equiv \underline{\check \tau } \nonumber \\
&&
\check g_\alpha \equiv \check g , \quad
\check g_{\bar \alpha }\equiv \underline{ \check g },\quad
\check G_\alpha \equiv \check G , \quad
\check G_{\bar \alpha }\equiv \underline{ \check G },
\end{eqnarray}
and so forth [see figure \ref{fig:notation}(a)]. Also, from Eq. \eqref{tausymm} we have $\check \tau = \check S \underline{\check\tau}^\dagger \underline{\check S}$.

\subsection{General Boundary Conditions for diffusive systems}

One main problem with boundary conditions for quasiclassical propagators is illustrated in figure \ref{fig:notation} (b) and (c). In previous treatments \cite{Millis88,naz,cottet} the starting point was a transfer matrix description, see figure  \ref{fig:notation} (b), which however required the elimination of so-called ``Drone amplitudes'', which are propagators that mix incoming with outgoing directions. Here, we will employ a scattering matrix description, see figure \ref{fig:notation} (c), which, on the other hand, requires a similar elimination of Drone amplitudes, this time being propagators mixing the two sides of the interface. However, for an impenetrable interface this latter problem does not arise, a fact we will exploit.

The strategy to derive the needed boundary conditions is to apply a three-step procedure. In the first step, the problem of an impenetrable interface with the auxiliary scattering matrix defined in Eq.~\eqref{SV0} is solved on each side of the interface  \cite{Eschrig03}. For this step, the ballistic solutions for the envelope functions for the Gor'kov propagators 
close to the interfaces should be expressed by the solutions $\check{G}$ of the Usadel equation.
In a second step, these ballistic solutions (auxiliary propagators) are used in order to find the full ballistic solutions for finite transmission by utilizing a $t$-matrix technique \cite{Cuevas01,Eschrig03,Eschrig04,Kopu04}.
In the third, and final, step the matrix current will be derived from the ballistic solutions, which then enters the boundary conditions for the Usadel equations. We will present explicit solutions for all three steps, such that the procedure describes effectively boundary conditions for the solutions of Usadel equations on either side of the interface.

We use for the auxiliary propagators the notation
$\check g_{0}^{o}$, $\check g_{0}^{i}$, 
$\underline {\check g}_{0}^{o}$ and $\underline {\check g}_{0}^{i}$, 
where
the upper index denotes the direction of the Fermi velocity. 
{\it Incoming} momenta (index $i$)
are those with a Fermi velocity pointing towards the interface, and {\it outgoing} momenta (index $o$) are those with a Fermi velocity pointing away from the interface.

\subsubsection{Solution for impenetrable interface:}
We solve first for the auxiliary ballistic propagators fulfilling the impenetrable boundary conditions
\begin{eqnarray}
\label{aux}
\check { g}_{0}^{o}= \check S\; \check { g}_{0}^{i} \; \check S^\dagger ,
\quad \underline{\check{ g}}_{0}^{o}= \underline{\check S}\; \underline{\check{ g}}_{0}^{i} \; \underline{\check S}^\dagger ,
\end{eqnarray}
implying matrix multiplication in the combined [Keldysh] $\times$ [particle-hole] $\times$ [combined scattering-channel and spin] space.
For diffusive banks, it is necessary to connect the ballistic propagators ${\check g}_{0}^{i,o}$ with the isotropic solutions of the Usadel equation, ${\check G}$. 
The ballistic propagators $\check{g}_{0}^{i,o}$ and $\underline{\check{g}}_{0}^{i,o}$, which characterize electronic correlations next to the
scattering barrier, depend on the electronic momentum. However, in the
diffusive case, impurity scattering leads to momentum isotropization away from
the scattering barrier. This process occurs in isotropization zones with a thickness corresponding to a few times the inelastic mean free path of the
materials, see figure \ref{fig:notation} (a). 
This scale is itself much smaller than the scale on which the
isotropic diffusive Green functions evolve in the bulk of the materials, in
the framework of the Usadel equations. Indeed, the Usadel equations involve
a superconducting coherence length, which
is typically much larger than the elastic mean free path. 
Therefore, in order to describe disordered hybrid
structures with Usadel equations, suitable boundary conditions should be
expressed in terms of the values of the isotropic Green functions $\check{G}$ 
and $\underline{\check{G}}$ right 
at the beginning of
the isotropization zones. To
obtain such boundary conditions from Eq. \eqref{aux}, it is necessary to express the
propagators $\check{g}_{0}^{i,o}$ and $\underline{\check{g}}_{0}^{i,o}$ in
terms of $\check{G}$ and $\underline{\check{G}}$. This can be done by studying
the spatial dependence of the Gor'kov Green functions (or full Green
functions without the quasiclassical approximation) in the isotropization
zones (see Refs. \cite{naz,cottet} for details). Using the fact that the dynamics of
electrons is dominated by impurity scattering in these zones, one can express
the Gor'kov Green functions in terms of $\check{g}_{0}^{i,o}$ ,
$\underline{\check{g}}_{0}^{i,o}$, $\check{G}$ and $\underline{\check{G}}$.
Then, an elimination of unphysical solutions imposes the conditions \cite{naz}
\numparts
\begin{eqnarray}
\label{naz1}
({\check G}-\i\pi {\check 1})\otimes (\check { g}_{0}^{i} + \i\pi {\check 1}) &=& {\check 0} 
,\quad
\label{naz2}
(\check { g}_{0}^{i} - \i\pi {\check 1}) \otimes ({\check G}+\i\pi {\check 1})= {\check 0} \\
\label{naz3}
({\check G}+\i\pi {\check 1})\otimes (\check { g}_{0}^{o} - \i\pi {\check 1}) &=& {\check 0} 
,\quad
\label{naz4}
(\check { g}_{0}^{o} + \i\pi {\check 1}) \otimes ({\check G}-\i\pi {\check 1})= {\check 0} 
\end{eqnarray}
\endnumparts
and similarly for $\underline{\check{G}}$ and $\underline{\check{g}}_{0}^{i,o}$.
From this one obtains the identity $
\frac{1}{2}\left\{ \check { g}_{0}^{i,o} , \check G\right\}_\otimes=-\pi^2 {\check 1}$ for the anticommutator $\left\{ \ldots \right\}$.
This allows to solve after some straightforward algebra for $\check { g}_{0}^{i,o}$, 
using Eq. \eqref{aux}, and using the abbreviations 
\begin{eqnarray}
\label{def1}
\check G'&=& \frac{1}{2\pi^2} \; (\check S^\dagger \check G \check S-\check G),
\label{def2}
\quad
\check G''= \frac{1}{2\pi^2} \; (\check S\check G\check S^\dagger -\check G) ,
\end{eqnarray}
(both are matrices depending via $\check S$ on the scattering channel index)
leading to  \cite{cottet}
\numparts
\begin{eqnarray}
\label{gid}
\check { g}_{0}^{i} -\i\pi {\check 1} = 
(1-\check G\otimes \check G')^{-1} \otimes
(\check G-\i\pi {\check 1} ) , \\
\label{god}
\check { g}_{0}^{o} +\i\pi {\check 1} = 
(1-\check G\otimes \check G'')^{-1} \otimes
(\check G+\i\pi {\check 1} ) 
\end{eqnarray}
(here and below the inverse is defined with respect to the $\otimes $-product),
which, using identities like $\check G' \otimes \check G'=-\frac{1}{2\pi^2}\left\{ \check G',\check G\right\}_\otimes $
(with $\left\{ A,B\right\}_\otimes \equiv A\otimes B+B\otimes A$),
alternatively can be written also as
\begin{eqnarray}
\label{gid1}
\check { g}_{0}^{i} +\i\pi {\check 1} =
 (\check G+\i\pi {\check 1} ) \otimes
(1-\check G'\otimes \check G)^{-1} , \\
\label{god1}
\check { g}_{0}^{o} -\i\pi {\check 1} =
 (\check G-\i\pi {\check 1} ) \otimes
(1-\check G''\otimes \check G)^{-1} .
\end{eqnarray}
\endnumparts
Similarly equations hold for $\underline{\check{G}}$ and $\underline{\check{g}}_{0}^{i,o}$ in terms of the scattering matrix $\underline{\check{S}}$.
Introducing these solutions into Eqs.~\eqref{naz1}-\eqref{naz4} shows readily that the latter are fulfilled.
We note that the relation
$\check g_{0}^{i,o}\otimes \check g_{0}^{i,o}=-\pi^2 \check 1$ follows from $\check G\otimes \check G=-\pi^2 \check 1$ and 
$\check S\check S^\dagger =
\check S^\dagger \check S= \check 1$.
It is also important to notice that whereas 
$\check G$ is proportional to the unit matrix in channel space due to their isotropic nature \cite{cottet}, $\check S $, and consequently $\check G'$, $\check G''$, and $\check g_{0}^{i,o}$, are in general non-trivial matrices in channel space. 
Eqs. \eqref{gid}-\eqref{god}, or alternatively \eqref{gid1}-\eqref{god1}, together with 
Eq. \eqref{def1}
determine uniquely $\check g_{0}^{i,o}$ in terms of the diffusive Green function $\check G$.
We can rewrite the difference $\check { g}_{0}^{o}-\check { g}_{0}^{i}$ in a more explicit manner, using the abbreviations $\check \delta' \equiv \check G\otimes \check G'$ and $\check \delta''\equiv \check G''\otimes \check G$, leading to
\begin{eqnarray}
\label{gomgi}
\check { g}_{0}^{o}-\check { g}_{0}^{i} = &&
(\check 1-\check \delta')^{-1}
\otimes \left[
(\check G-\i\pi {\check 1} ) \otimes \check \delta'' 
- \check \delta' \otimes (\check G-\i\pi {\check 1} ) 
\right] \otimes
(\check 1-\check \delta'')^{-1} .
\end{eqnarray}

\subsubsection{Solution for finite transmission:}
The second step follows Refs.~ \cite{Eschrig03,Eschrig04}.
Once the auxiliary propagators are obtained, the full propagators 
can be obtained directly, without further solving the transport equation, in the following way. 
We solve {\it $t$-matrix equations} resulting from the transmission parameters $\check \tau$,
for incoming and outgoing directions, which according to a procedure analogous to the one discussed in Ref. \cite{Cuevas96,Cuevas01} take the form,
\begin{eqnarray}
\label{ti}
\check { t}^{i}&=&
\underline{\check \tau}^\dagger \; \underline{\check g}_{0}^{o} \;
\underline{\check \tau}
\otimes \left( \check 1 +
\check { g}_{0}^{i} \otimes \check { t}^{i}\right), 
\quad
\label{to}
\check { t}^{o}=
\check \tau\; \underline{\check  g}_{0}^{i} \;
\check \tau^\dagger
\otimes \left( \check 1 +
\check { g}_{0}^{o} \otimes \check { t}^{o}\right).
\end{eqnarray}
Using the symmetry Eq.~\eqref{tausymm},
the $t$-matrices for incoming and outgoing directions can be related 
through
\begin{eqnarray}
\label{tsym}
\check { t}^{o}&=&
\hat S\; 
\check { t}^{i} \;
\hat S^\dagger .
\end{eqnarray}
Using the short notation 
\begin{eqnarray}
\label{def01}
\check { g}_{1}^{o} &\equiv &
\check \tau\; \underline{\check g}_{0}^{i} \;
\check \tau^\dagger, \qquad
\check { g}_{1}^{i} 
\equiv
\underline{\check \tau}^\dagger \; \underline{\check g}_{0}^{o} \;
\underline{\check \tau},
\end{eqnarray}
we solve formally Eqs.~\eqref{to} for $\check t^{i,o}$:
\begin{eqnarray}
\label{tmatrix}
\check { t}^{i,o}&=& 
\left(1- \check { g}_{1}^{i,o} \otimes \check { g}_{0}^{i,o} \right)^{-1}
\otimes \check { g}_{1}^{i,o}.
\end{eqnarray}

The {\it full propagators}, fulfilling
the desired boundary conditions at the interface, can now be easily calculated.
For incoming and outgoing directions they are obtained from \cite{Eschrig03,Kopu04}
\numparts
\begin{eqnarray}
\label{gi}
\check { g}^{i}&=& 
\check { g}_{0}^{i} 
+ \left( \check { g}_{0}^{i} + \i\pi\check 1 \right) \otimes
\check { t}^{i} \otimes \left(\check { g}_{0}^{i}- \i\pi\check 1\right), \quad \\
\check { g}^{o}&=&
\check { g}_{0}^{o} +
\left(\check { g}_{0}^{o} - \i\pi\check 1\right) \otimes
 \check { t}^{o} \otimes \left(\check { g}_{0}^{o} + \i\pi\check 1\right). \quad
\label{go}
\end{eqnarray}
Noticing that 
$\left( \check { g}_{0}^{i,o} + \i\pi\check 1 \right) \otimes \left(\check { g}_{0}^{i,o}- \i\pi\check 1\right)=\check 0$, and 
$\left( \check { g}_{0}^{i,o} - \i\pi\check 1 \right) \otimes \left(\check { g}_{0}^{i,o}+ \i\pi\check 1\right)=\check 0$, as well as identities like ${\check { g}}_{0}^{i,o} \otimes ({\check { g}}_{0}^{i,o} +\i\pi {\check 1})= \i\pi {\check 1} \otimes ({\check { g}}_{0}^{i,o}+\i\pi {\check 1})$ etc,
it is obvious that the normalization
$\check { g}^{i,o}\otimes \check { g}^{i,o}=-\pi^2 \check 1$ holds.
Using the same identities, we obtain the alternative to Eqs. \eqref{gi}-\eqref{go} expressions
\begin{eqnarray}
\label{gi1}
\check { g}^{i}
&=& \check { g}_{0}^{i}+(\check { g}_{0}^{i}+\i\pi \check 1) \otimes [ \check { t}^{i} , \check { g}_{0}^{i} ]_\otimes 
=\check { g}_{0}^{i}- [ \check { t}^{i} , \check { g}_{0}^{i} ]_\otimes  \otimes
(\check { g}_{0}^{i}-\i\pi \check 1) ,\\
\label{go1}
\check { g}^{o}&= &
\check { g}_{0}^{o}+
(\check { g}_{0}^{o}-\i\pi \check 1) \otimes [ \check { t}^{o} , \check { g}_{0}^{o} ]_\otimes 
=\check { g}_{0}^{o}- [ \check { t}^{o} , \check { g}_{0}^{o} ]_\otimes  \otimes
(\check { g}_{0}^{o}+\i\pi \check 1) .
\end{eqnarray}
\endnumparts
Equations \eqref{gi}-\eqref{go}, or alternatively, \eqref{gi1}-\eqref{go1}, in conjunction with Eqs. \eqref{def01}-\eqref{tmatrix}, solve the problem of finding the ballistic solutions for finite transmission. We are now ready for the last step, to relate these solutions to the matrix current which 
enters in the expression for boundary conditions for $\check G$ and $\underline{\check G}$.

\subsubsection{Matrix current and boundary conditions for diffusive propagators:}
We now turn to the third, final, step.
As shown in Refs. \cite{naz,cottet}, the boundary conditions for quasiclassical isotropic
Green functions can be obtained from the conservation of the matrix current
$\mathcal{I}$ in the isotropization zones surrounding the scattering barrier.
This quantity contains physical information on the flows of charge, spin and
electron-hole coherence in a structure. We refer the reader to Refs. \cite{naz,cottet} for
the general definition of $\mathcal{I}$ in terms of the Gor'kov Green
functions. Using this definition, one can verify that $\mathcal{I}$ is
spatially conserved along the entire isotropization zones. Then, one can
express $\mathcal{I}$ next to the scattering barrier in terms of the
propagators $\check{g}^{i,o}$ and $\underline{\check{g}}^{i,o}$, and
at the beginning of the isotropization zones in terms of $\check{G}$ and
$\underline{\check{G}}$, see Fig. \ref{fig:notation} (a). The conservation of the matrix current provides an
equality between the two expressions. Since $\check{g}^{i,o}$ can be
expressed in terms of $\check{g}_0^{i,o}$ and $\underline{\check{g}}_0^{i,o}$, and these in terms of
the $\check{G}$ and $\underline{\check{G}}$, this gives
the desired boundary conditions.
Following Ref.~ \cite{Kopu04}, after some straightforward algebra we obtain
\begin{eqnarray}
\label{comm1}
&&[ \check { t}^{o} , \check { g}_{0}^{o} ]_\otimes = 
\left(1- \check { g}_{1}^{o} \otimes \check { g}_{0}^{o} \right)^{-1}
\left[ \check { g}_{1}^{o}, \check { g}_{0}^{o} \right]_\otimes
\left(1- \check { g}_{0}^{o} \otimes \check { g}_{1}^{o} \right)^{-1} .\qquad
\end{eqnarray}
Using 
relations \eqref{aux} and \eqref{tsym} above, we find
\begin{eqnarray}
\check { g}^{i}&=&
\check S^\dagger \; \left[
\check { g}_{0}^{o} +
\left(\check { g}_{0}^{o} + \i\pi\check 1\right) \otimes
 \check { t}^{o} \otimes \left(\check { g}_{0}^{o} - \i\pi\check 1\right)
\right] \check S, 
\label{eq8}
\end{eqnarray}
which allows to derive the following relation
\begin{eqnarray}
\label{comm}
\check{\cal I}'\equiv
\check { g}^{o}- \check S\check { g}^{i} \check S^\dagger
&=& -2\pi \i[ \check { t}^{o} , \check { g}_{0}^{o} ]_\otimes .
\end{eqnarray}
For calculating the charge current density in a given structure, it is sufficient to know $\check{\cal I}'$, because the matrices $\check S$ and $\check S^\dagger $ drop out of the trace as they commute with the $\hat \tau_3$ matrix in particle-hole space.

Finally we relate the obtained propagators $\check g^{i,o}$ to the matrix current ${\cal I}$,
\begin{eqnarray}
\label{matrixcurrent}
\check{\cal I}\equiv
\check { g}^{o}- \check { g}^{i} \equiv
\check{\cal I}'+ \check{\cal I}''
\end{eqnarray}
with
\begin{eqnarray}
\label{matrixcurrent2}
\check{\cal I}''\equiv
\check S\check { g}^{i} \check S^\dagger -
\check { g}^{i}  .
\end{eqnarray}

We remind the reader here that $\check{\cal I}$ has a matrix structure in Keldysh space, in particle-hole space, and in combined scattering-channel and spin space.
In terms of $\check{\cal I}$ the boundary condition results then from 
Eq.~\eqref{relation} and from the matrix current conservation in the isotropization regions \cite{naz} 
\begin{eqnarray}
\label{gl_naz}
{\cal G}_q\sum_{n=1}^{\cal N}
\frac{\check{\cal I}_{nn} }{\i\pi }
=
-\frac{\sigma {\cal A}}{\pi^2}\check{G}\otimes \frac{d}{dz}\check{G},
\end{eqnarray}
where $z$ is the coordinate along the interface normal 
({\it away from} the interface), $n$ is a scattering channel index (${\cal N}$ channels, spin-degenerate channels count as one),
$\sigma=e^2N_{{\rm F}} D$ refers to the conductivity per spin,
${\cal A}$ is the surface area of the contact, 
and ${\cal G}_q$ is the quantum of conductance, 
${\cal G}_q= e^2/h$. 
The number of scattering channels is expressed in terms of the projection of the Fermi surfaces on the contact plane, $A_{F,z}$, by ${\cal N}= A_{F,z}{\cal A}/(2\pi)^2$. For isotropic Fermi surfaces $A_{F,z}=\pi k_F^2$.
In general,
\begin{eqnarray}
\frac{1}{\cal A} \sum_{n=1}^{\cal N} \ldots = \int_{A_{F,z}} \frac{d^2 k_{||}}{(2\pi)^2} \ldots  ,
\end{eqnarray}
where $\hbar {\bf k}_{||}$ is the momentum component parallel to the interface.

\section{Special Cases}
\subsection{Spin-scalar and channel-diagonal case}
The transition to the diffusive Green functions is trivial for the case of $\hat S=\hat 1$, as then $\check g_{0}^{i}=\check g_{0}^{o}=\check G$. 
If we start from Eq.~\eqref{comm1} in conjunction with \eqref{def01}, we obtain in the case of a spin-scalar and channel-diagonal matrix $\hat \tau_{nn}$ 
with the notation $\check G = -\i\pi \check {\bf G}$ 
\numparts
\begin{eqnarray}
\frac{2\sum_n\check{\cal I}_{nn}}{\i\pi }= 
\sum_n\frac{
4{\cal T}_n[\underline{\check{\bf G}},\check{\bf G}]
}{
4 +{\cal T}_n\left(\{\underline{\check{\bf G}},\check{\bf G}\}-2\right)
}
=\frac{2\sigma {\cal A}}{{\cal G}_q} \check{\bf G} \otimes \frac{d}{dz}\check{\bf G}
\end{eqnarray}
with $\sigma=e^2N_FD$ and
\begin{eqnarray}
{\cal T}_n=\frac{4\pi^2 |\tau_{nn}|^2}{\left(1+\pi^2|\tau_{nn}|^2\right)^2} .
\end{eqnarray}
\endnumparts
This reproduces Nazarov's boundary condition \cite{naz,Kopu04}.

\subsection{Case for interface between superconductor and ferromagnetic insulator}
For the case of zero transmission, $\check \tau\equiv \check 0$, we can find a
closed solution if we assume that we can find a spin-diagonal basis for all reflection channels. 
For a channel-diagonal scattering matrix
we write $\check S_{nn}=e^{\i\varphi_n} e^{\i\frac{\vartheta_n}{2}\check \kappa }$
with $\check \kappa = \mbox{diag}\left\{ \vec{m}\vec{\sigma},\vec{m}\vec{\sigma}^\ast \right\}$, where $\vec{m}^2=1$ (leading to $\check \kappa^2=1$).
In this case we have $\check g^{i,o}=\check g^{i,o}_0$. We
use Eq. \eqref{gomgi}, which straightforwardly leads to
\begin{eqnarray}
\frac{2\sum_n\check {\cal I}_{nn}}{\i\pi}&= &
\sum_n
\left[ \check 1-\frac{\i\sin \vartheta_n}{4} (\check {\bf G} \check \kappa \check {\bf G}-\check \kappa ) +\frac{\sin^2 \frac{\vartheta_n}{2}}{2} (\check {\bf G} \check \kappa \check {\bf G} \check \kappa - \check 1) \right]^{-1}
\nonumber \\
&&\qquad \times \left\{ -\i\sin \vartheta_n [\check \kappa,\check {\bf G}] + \sin^2 \frac{\vartheta_n}{2}[\check \kappa \check {\bf G} \check \kappa, \check {\bf G}] \right\} \nonumber \\
&&\times \left[ \check 1-\frac{\i\sin \vartheta_n}{4} (\check {\bf G} \check \kappa \check {\bf G}-\check \kappa ) +\frac{\sin^2 \frac{\vartheta_n}{2}}{2} (\check \kappa \check {\bf G} \check \kappa \check {\bf G} - \check 1) \right]^{-1} 
\label{FI}
\end{eqnarray}
(where we remind that $\check {\bf G}^2 =\check 1$). Note that $\varphi_n $ drops out, only the spin mixing angle $\vartheta_n $ matters. 
Eq. \eqref{FI} generalizes the results of Ref. \cite{cottet} to arbitrary spin-dependent reflection phases. Further below we will give a physical interpretation of the leading order terms arising in an expansion for small $\vartheta_n$.

\subsection{Exact series expansions}
\label{series}
We now provide explicit series expansions for all quantities which will be useful for deriving formulas for various limiting cases. We start with writing the scattering matrix as $\check S=e^{\i\check K}$ with hermitian $\check K$ due to unitarity of $\check S$, \ie $\check K=\check K^\dagger $. Then we use an expansion formula for Lie brackets 
in order to obtain the series expansion
\begin{eqnarray}
\label{Lie}
\check S^\dagger \check G \check S = e^{-\i\check K} \check G e^{\i\check K} = \sum_{m=0}^\infty \frac{(-\i)^m}{m!} \left[ \check K \stackrel{m}{,} \check G\right]
\end{eqnarray}
with the definitions $\left[ \check K \stackrel{m}{,} \check G\right] = \left[ \check K , \left[ \check K\stackrel{m-1}{,} \check G\right]\right]$ and $\left[ \check K \stackrel{0}{,} \check G\right]=\check G$. With this we obtain from 
Eq. \eqref{def1}
\begin{eqnarray}
\check G' &=&\frac{1}{2\pi^2}
\sum_{m=1}^\infty \frac{(-\i)^m}{m!} \left[ \check K \stackrel{m}{,} \check G\right] ,
\qquad
\check G'' =\frac{1}{2\pi^2}
\sum_{m=1}^\infty \frac{\i^m}{m!} \left[ \check K \stackrel{m}{,} \check G\right] ,
\end{eqnarray}
which are very useful if $\check K$ has a small pre-factor.
Note also the identity $\check G \otimes \left[ \check K , \check G\right] \otimes \check G= \pi^2 \left[ \check K , \check G\right]$.
Furthermore, from Eqs. \eqref{gid1}-\eqref{god1} we find
\numparts
\begin{eqnarray}
\check g_0^i &=& \check G+ (\check G+\i\pi \check 1) \otimes \sum_{l=1}^\infty
(\check G' \otimes \check G)^l \\
\check g_0^o &=& \check G+ (\check G-\i\pi \check 1) \otimes \sum_{l=1}^\infty
(\check G''\otimes \check G)^l.
\end{eqnarray}
\endnumparts
From Eq. \eqref{comm1}, and using Eqs. \eqref{aux}, \eqref{tsym}, we derive
\numparts
\begin{eqnarray}
\left[ \check t^o,\check g^o_0\right]_\otimes &=&
\sum_{k,n=0}^\infty (\check g_1^o \otimes \check g_0^o)^k \otimes\left[ \check g_1^o, \check g_0^o\right]_\otimes \otimes (\check g_0^o \otimes \check g_1^o)^n,\quad \\
\left[ \check t^i,\check g^i_0\right]_\otimes &=&
\sum_{k,n=0}^\infty (\check g_1^i \otimes \check g_0^i)^k \otimes\left[ \check g_1^i, \check g_0^i\right]_\otimes \otimes (\check g_0^i \otimes \check g_1^i)^n,
\end{eqnarray}
\endnumparts
which is useful if the transmission amplitudes $\check \tau $ entering into $\check g_1^{i,o} $ are small.
Finally, we obtain from Eqs. \eqref{comm} and
\eqref{matrixcurrent2} 
\begin{eqnarray}
\check{\cal I}'=-2\pi \i \left[ \check t^o,\check g^o_0\right]_\otimes,\quad
\check {\cal I}'' &=&
\sum_{m=1}^\infty \frac{\i^m}{m!} \left[ \check K \stackrel{m}{,} \check g^i\right] .
\end{eqnarray}
Here, $\check g^i$ is obtained from
\begin{eqnarray}
\label{seriesgi}
\check g^i+\i\pi \check 1= (\check G+\i\pi \check 1) \otimes
\sum_{l=0}^\infty (\check G'\otimes \check G)^l \otimes \left(\check 1+\left[ \check t^i,\check g_0^i\right]_\otimes\right)  . \quad
\end{eqnarray}

\subsection{Boundary condition for spin-polarized surface to third order in spin-mixing angles}

We first treat the case when $\check t^{i,o}\equiv \check 0$, for example the case where one side of the junction is a ferromagnetic insulator (FI). Then
\begin{eqnarray}
\label{expansion}
\check{\cal I} &=&
\sum_{m=1}^\infty \frac{\i^{m}}{m!} \left[ \check K \stackrel{m}{,} \check G \right]
+\sum_{m,l=1}^\infty \frac{\i^{m}}{m!} \left[ \check K \stackrel{m}{,} 
(\check G+\i\pi \check 1) \otimes (\check G' \otimes \check G)^l 
\right] .
\end{eqnarray}
To third order we have $\check{\cal I}=\check{\cal I}^{(1)} +\check{\cal I}^{(2)}+\check{\cal I}^{(3)}$, 
and the derivation in \ref{thirdorder} leads to
\numparts
\begin{eqnarray}
\label{I3}
&&\check{\cal I}^{(1)} = \i\left[ \check K , \check G\right],\qquad
\check{\cal I}^{(2)} = 
-\frac{\i}{2\pi} \left[ \check K  \check G \check K ,\check G \right]_\otimes \quad \\
\label{three}
&&\check{\cal I}^{(3)} = 
-\frac{\i}{24}\left[ \check K \stackrel{3}{,} \check G\right] 
-\frac{\i}{8\pi^2 }\left[ \check K , \check G \otimes \left[ \check K \stackrel{2}{,} \check G\right] \otimes \check G \right] .  \qquad
\end{eqnarray}
\endnumparts
For the special case of channel diagonal $\check K_{nn}=\frac{\vartheta_n}{2} \check \kappa $ with $\check \kappa^2=\check 1$,  which follows also from directly expanding Eq. \eqref{FI},
we reproduce the results from Ref.  \cite{cottet} ($\check G=-i\pi \check{\bf G}$),
\numparts
\begin{eqnarray}
&&\frac{2\sum_n\check{\cal I}_{nn}^{(1)}}{\i\pi} = -\i\left(\mbox{$\sum_n$}\vartheta_n\right) \left[\check \kappa,\check{\bf G}\right] ,\quad
\frac{2\sum_n\check{\cal I}_{nn}^{(2)}}{\i\pi} = \frac{\sum_n \vartheta_n^2}{4} \left[\check \kappa\check {\bf G}\check \kappa ,\check {\bf G} \right]_\otimes \\
&&\frac{2\sum_n\check{\cal I}_{nn}^{(3)}}{\i\pi} = -\i \frac{\sum_n\vartheta_n^3}{16} \left(\frac{1}{3} \left[\check \kappa,\check {\bf G}\right] - \left[\check \kappa\check {\bf G}\check \kappa \otimes \check {\bf G}\check \kappa, \check {\bf G} \right]_\otimes
\right)
. \quad
\end{eqnarray}
\endnumparts
Note that the first order term $\sim[\check \kappa,\check {\bf G}]$ accounts for the effective exchange field induced inside the superconductor by the spin-mixing, whereas the term $\sim[\check \kappa \check {\bf G} \check \kappa,\check {\bf G}]$ produces a pair breaking effect similar to that of  paramagnetic impurities \cite{Abrikosov60}. This second term occurs only at second order in $\vartheta_n$ because it requires multiple scattering at the S/FI interface, which together with random scattering in the diffusive superconductor leads to a magnetic disorder effect. 

\subsection{Boundary condition for spin-polarized interface to second order in spin-mixing angles and transmission probability}

We now allow for finite transmission, and
concentrate on the matrix current to second order in the quantities $\check K$, $\underline{\check K}$, and $\check g^{i,o}_1$.
We need to take care of the scattering phases during transmission events. For this, we define 
\begin{equation}
\check \tau=\check S^{\frac{1}{2}} \check \tau_0 \underline{\check S}^{\frac{1}{2}}
,\quad \underline{\check \tau}= \underline{\check S}^{\frac{1}{2}} \underline{\check \tau}_0 \check S^{\frac{1}{2}} .
\end{equation}
We note that Eq. \eqref{tausymm}, or $\check \tau=\check S \underline{\check \tau}^\dagger \underline{\check S}$, results into
\begin{equation}
\check \tau_0=\underline{\check \tau}^\dagger_0. 
\end{equation}
Thus, the $\check \tau_0$ and $\underline{\check \tau}_0$ are the appropriate transmission amplitudes, with transmission spin-mixing phases removed. 
We further define
\begin{equation}
\check G_1\equiv \tau_0 \underline{\check G} \tau_0^\dagger .
\end{equation}
We expand $\check \tau $ up to first order in $\check K$ and $\underline{\check K}$,
\begin{equation}
\check \tau = \check \tau_0+ \frac{\i}{2} \left( \check K \check \tau_0+ \check \tau_0 \underline{\check K} \right) + \ldots ,
\end{equation}
and obtain
$\check{\cal I}=\check{\cal I}^{(1)} +\check{\cal I}^{(2)}$ from a systematic expansion to second order in $\check K$, $\underline{\check K}$, and $\check G_1$, as shown in \ref{secondorder}, leading to one of the main results of this paper
\numparts
\begin{eqnarray}
\label{mainBC1}
\check{\cal I}^{(1)}&=&-2\pi \i 
\left[ \check G_1 ,\check G \right]_\otimes
+\i \left[ \check K ,\check G \right] , 
\\
\label{mainBC2}
\check{\cal I}^{(2)}&=&-2\pi \i
\left[ \check G_1 \otimes \check G \otimes \check G_1 ,\check G \right]_\otimes 
-\frac{\i}{2\pi} \left[ \check K \check G\check K,\check G\right]_\otimes
\nonumber \\
&&+\i\left[ \check G_1 \otimes \check G \check K+\check K \check G \otimes \check G_1
+\check \tau_0 \underline{\check G} \otimes \left[ \underline{\check K} ,\underline{\check G} \right] \check \tau_0^\dagger
, \check G\right]_\otimes  .
\end{eqnarray}
\endnumparts
These relations generalize the results of Ref.  \cite{cottet} for the case of arbitrary spin polarization, and are valid even when
$\check K$, $\underline{\check K}$ and $\tau $ have different spin quantization axes, i.e. cannot be diagonalized simultaneously.

Using the notation $\check G = -\i\pi \check {\bf G} $ and
$2\pi \check \tau_0 = \check T$, we can rewrite the result in leading order in the quantities $\check K$, $\underline{\check K}$, and the transmission probability
($\sim\check T\check T^\dagger$) as
\numparts
\begin{eqnarray}
\label{mainBCa1}
&&\frac{2\check{\cal I}^{(1)}}{\i\pi}=
\left[ 
\check T\; \underline{\check {\bf G}} \; \check T^\dagger -2\i\check K
,\check {\bf G} \right]_\otimes 
,
\end{eqnarray}
and for the next to leading order
\begin{eqnarray}
\frac{2\check{\cal I}^{(2)}}{\i\pi}&=&
-\frac{1}{4}
\left[ \check T\; \underline{\check {\bf G}} \; \check T^\dagger \otimes
\check {\bf G} \otimes \check T\; \underline{\check {\bf G}} \; \check T^\dagger
,\check {\bf G} \right]_\otimes 
+\left[ \check K \check {\bf G} \check K , \check {\bf G} \right]_\otimes 
\nonumber \\
\label{mainBCa2}
& +& \frac{\i}{2} \left[ \check T\; \underline{\check {\bf G}} \; \check T^\dagger \otimes \check {\bf G} \check K+ \check K  \check {\bf G} \otimes \check T\; \underline{\check {\bf G}} \; \check T^\dagger 
+\check T \underline{\check {\bf G} } \otimes \left[ \underline{\check K}, \underline{\check {\bf G} } \right] \check T^\dagger
, \check {\bf G}  \right]_\otimes  .
\end{eqnarray}
\endnumparts
These equations are still fully general with respect to the magnetic (spin) structure, and allow for channel off-diagonal scattering as well as different numbers of channels on the two sides of the interface. Note that $\check T$, $\check K$, and $\underline{\check K}$ are matrices in channel space, whereas $\check {\bf G}$ and $\underline{\check{\bf G}}$ are proportional to the unit matrix in channel space. Whereas $\check K$, and $\underline{\check K}$ are square matrices, $\check T$ in general can be a rectangular matrix (when the number of channels on the two sides of the interface differ).

\subsection{Boundary conditions for channel-independent spin quantization direction}

As an application, we assume next
that each of the quantities $\check K$, $\underline{\check K}$, and $\check \tau $ 
can be spin-diagonalized simultaneously for all channels, 
with spin quantization directions $\vec{m}'$, $\underline{\vec{m}}'$, and $\vec{m}$  for $\check K$, $\underline{\check K}$, or $\check \tau $, respectively. 
We also use that $\check {\bf G}$ and $\underline{\check {\bf G}}$ are proportional to the unit matrix in channel space, as they are isotropic \cite{cottet}, and we assume that the number of channels on both sides of the interface are equal.
We define
\numparts
\begin{eqnarray}
&&\mathbb{T}_{0,nl}\; \check 1+\mathbb{T}_{1,nl} \; \vec{m}\cdot \vec{\check\sigma} =\check T_{nl} , 
\\
&&\varphi_{nn'} \; \check 1+\frac{1}{2}\vartheta_{nn'} \; \vec{m}'\cdot \vec{\check\sigma} =
\check K_{nn'}, \quad
\underline{\varphi}_{ll'}\; \check 1+\frac{1}{2}\underline{\vartheta}_{ll'} \; \underline{\vec{m}}' \cdot\vec{\check\sigma} =
\underline{\check K}_{ll'}, 
\\
&&\vec{\check\sigma} = \vec{\hat\sigma}\check 1,\quad
\vec{\hat\sigma}=\left( \begin{array}{cc} \vec{\sigma}&0\\0& \vec{\sigma}^\ast \end{array} \right)_{\! \rm ph}, \check \kappa \equiv \vec{m}\cdot \vec{\check\sigma}, \quad
\check\kappa' \equiv \vec{m}' \cdot \vec{\check\sigma}, \quad
\underline{\check\kappa}' \equiv \underline{\vec{m}}' \cdot\vec{\check\sigma} 
\end{eqnarray}
\endnumparts
with $\vec{m}^2=(\vec{m}')^2=(\underline{\vec{m}}')^2=1$, i.e. $\check \kappa^2=(\check \kappa')^2=(\underline{\check\kappa}')^2=\check 1$,
and introduce the transmission probability ${\cal T}_{nl}$ and the spin polarization
${\cal P}_{nl}$ as
\begin{eqnarray}
&&{\cal T}_{nl}\left(\check 1+{\cal P}_{nl} \vec{m}\vec{\check\sigma} \right)=
\check T_{nl} [\check T_{nl}]^\dagger.
\end{eqnarray}
We write for $\mathbb{T}_{0,nl}$ and $\mathbb{T}_{1,nl}$, allowing for some spin-scalar phases $\psi_{nl}$,
\begin{eqnarray}
\label{Tfactors}
\mathbb{T}_{0,nl}^2= \frac{{\cal T}_{nl}}{2} \left[ 1+\sqrt{1-{\cal P}_{nl}^2} \right]e^{2\i\psi_{nl}},\;
\mathbb{T}_{1,nl}^2= \frac{{\cal T}_{nl}}{2} \left[ 1-\sqrt{1-{\cal P}_{nl}^2} \right]e^{2\i\psi_{nl}}.
\end{eqnarray}
We will average over all spin-scalar phases $\psi_{nl}$ of the transmission amplitudes as there are usually many scattering channels in an area comparable with the superconducting coherence length squared. This filters out all the terms
in Eqs. \eqref{mainBCa1}-\eqref{mainBCa2} where these scalar scattering phases cancel.

For a magnetic system, in linear order in ${\cal T}_{nl}$ and $\vartheta_{nn'}$ we obtain 
\begin{eqnarray}
I^{(1)}\equiv\frac{2{\cal G}_q\sum_n\check{\cal I}^{(1)}_{nn}}{\i\pi}
&=&{\cal G}_q\mbox{$\sum_{nl}$} \left[(\mathbb{T}_{0,nl}\check 1+\mathbb{T}_{1,nl}\check{\kappa})\underline{\check{\bf G}} (\mathbb{T}_{0,nl}^\ast\check 1+\mathbb{T}_{1,nl}^\ast\check{\kappa}), \check{\bf G}\right] 
\nonumber \\ 
&&-{\cal G}_q\mbox{$\sum_{n}$} \i \vartheta_{nn} 
\left[\check \kappa', \check{\bf G}\right],
\end{eqnarray}
where ${\cal G}_q=e^2/h$ is the conductance quantum.
After multiplying out we obtain the set of boundary conditions
\numparts
\begin{eqnarray}
2I^{(1)}&=&
\left[{\cal G}^0\underline{\check{\bf G}}+
{\cal G}^{\rm MR}\left\{\check{\kappa},\underline{\check{\bf G}}\right\} 
+{\cal G}^{1}\check{\kappa}\underline{\check{\bf G}}\check{\kappa}-\i{\cal G}^{\phi}_{}\check{\kappa}',\check{\bf G}\right]_\otimes
\label{newBC}
\end{eqnarray}
with
\begin{eqnarray}
{\cal G}^{0}  &=& {\cal G}_q
\mbox{$\sum\nolimits_{nl}$}
{\cal T}_{nl}\label{GT}\left( 1 + \sqrt{1-{\cal P}_{nl}^2}\right)\\
{\cal G}^{1}    &=&{\cal G}_q
\mbox{$\sum\nolimits_{nl}$}
{\cal T}_{nl}\left( 1- \sqrt{1-{\cal P}_{nl}^2}\right)\label{GMR2}\\
{\cal G}^{\rm MR}    &=&{\cal G}_q
\mbox{$\sum\nolimits_{nl}$}
{\cal T}_{nl}{\cal P}_{nl}\label{GMR}, \qquad
{\cal G}^{\phi}_{}    =2{\cal G}_q
\mbox{$\sum\nolimits_{n}$}
\vartheta_{nn}\label{Gfi}
\end{eqnarray}
\endnumparts
For $\kappa=\kappa'$ and the assumption of a channel-diagonal scattering matrix ($n=l$) this also provides the derivation of the boundary conditions used for Ref.~ \cite{Machon1}.
We now proceed to the second order terms: 
\numparts
\begin{eqnarray}
\label{newBC2}
2I^{(2)}&=& -2I_4 
+{\cal G}^{\phi}_2 \left[\check \kappa' \check{\bf G} \check \kappa' , \check{\bf G}  \right]_\otimes
+\i\left[ 
\check {\bf M}_{\chi,\underline{\chi}}^0
+\check {\bf M}_{\chi,\underline{\chi}}^1
+\check {\bf M}_{\chi,\underline{\chi}}^{\rm MR}
, \check{\bf G}  \right]_\otimes \\
\check {\bf M}_{\chi,\underline{\chi}}^0&=&
{\cal G}_{\chi}^0\; \left(\underline{\check{\bf G}} \otimes \check{\bf G} \check \kappa' + \check \kappa'  \check{\bf G}\otimes \underline{\check{\bf G}} \right)
+ {\cal G}_{\underline\chi}^0\; \underline{\check{\bf G}} \otimes \left[\underline{\check \kappa}',\underline{\check{\bf G}}\right] 
\nonumber \\
\check {\bf M}_{\chi,\underline{\chi}}^1&=&
 {\cal G}_{\chi}^1 \; \left(\check \kappa\underline{\check{\bf G}}\check \kappa \otimes \check{\bf G} \check \kappa' + \check \kappa'  \check{\bf G}\otimes \check \kappa\underline{\check{\bf G}}\check \kappa \right)
+{\cal G}_{\underline \chi}^1 \; \check \kappa \underline{\check{\bf G}} \otimes \left[\underline{\check \kappa}',\underline{\check{\bf G}}\right]\check \kappa 
\nonumber \\
\check {\bf M}_{\chi,\underline{\chi}}^{\rm MR}&=&
 {\cal G}_{\chi}^{\rm MR} \left(\left\{\check \kappa,\underline{\check{\bf G}} \right\} \otimes \check{\bf G} \check \kappa' + \check \kappa'  \check{\bf G}\otimes \left\{ \check \kappa,\underline{\check{\bf G}}\right\} \right)
+{\cal G}_{\underline \chi}^{\rm MR} \left\{ \check \kappa,\underline{\check{\bf G}} \otimes \left[\underline{\check \kappa}', \underline{\check{\bf G}}\right]\right\}
\nonumber
\end{eqnarray}
where $I_4$ denotes a cumbersome expression in fourth order of the transmission amplitudes, which we do not write down here explicitly (see \ref{I4}).
We have used the abbreviations
\begin{eqnarray}
\label{Gchi1}
{\cal G}_\chi^{0}  &=& \frac{1}{4}{\cal G}_q
\mbox{$\sum\nolimits_{nl}$}
\vartheta_{nn}{\cal T}_{nl}\left( 1 + \sqrt{1-{\cal P}_{nl}^2}\right)\\
{\cal G}_\chi^{1}  &  =&\frac{1}{4} {\cal G}_q
\mbox{$\sum\nolimits_{nl}$}
\vartheta_{nn} {\cal T}_{nl}\left( 1- \sqrt{1-{\cal P}_{nl}^2}\right)\\
{\cal G}_\chi^{\rm MR}    &=&\frac{1}{4} {\cal G}_q
\mbox{$\sum\nolimits_{nl}$}
\vartheta_{nn}{\cal T}_{nl}{\cal P}_{nl} , \qquad
{\cal G}^{\phi}_2    =\frac{1}{2}{\cal G}_q
\mbox{$\sum\nolimits_{nn'}$}
\vartheta_{nn'}^2
\label{Gchi2}
\end{eqnarray}
\endnumparts
and ${\cal G}_{\underline\chi}^0$, ${\cal G}_{\underline\chi}^1$, ${\cal G}_{\underline\chi}^{\rm MR}$ are defined as ${\cal G}_\chi^{0}$, ${\cal G}_\chi^{1}$, and $ {\cal G}_{\chi}^{\rm MR} $ with $\vartheta_{nn}$ replaced by $\underline{\vartheta}_{ll}$.
Note that $\varphi_{nn'}$ and $\underline{\varphi}_{ll'}$ do not appear in these expressions, in accordance with the intuitive notion that scalar scattering phases should drop out in the quasiclassical limit, which operates with envelope functions only.

The case for only channel-conserving scattering (channel-diagonal problem)
follows by taking in Eqs. \eqref{GT}-\eqref{Gfi} and
\eqref{Gchi1}-\eqref{Gchi2} only the terms with $n=l$. All other formulas \eqref{newBC}, \eqref{newBC2} remain unchanged.
This case is treated in Ref. \cite{cottet} to linear order in ${\cal P}_{nn}$, and our formulas reduce to these results for the considered limit. Note that for this case all spin-scalar phases cancel automatically and no averaging procedure over these phases is necessary.

\section{Application for diffusive superconductor/half metal heterostructure}

\begin{figure}[t!]
\centering{
\includegraphics[width=0.6\linewidth]{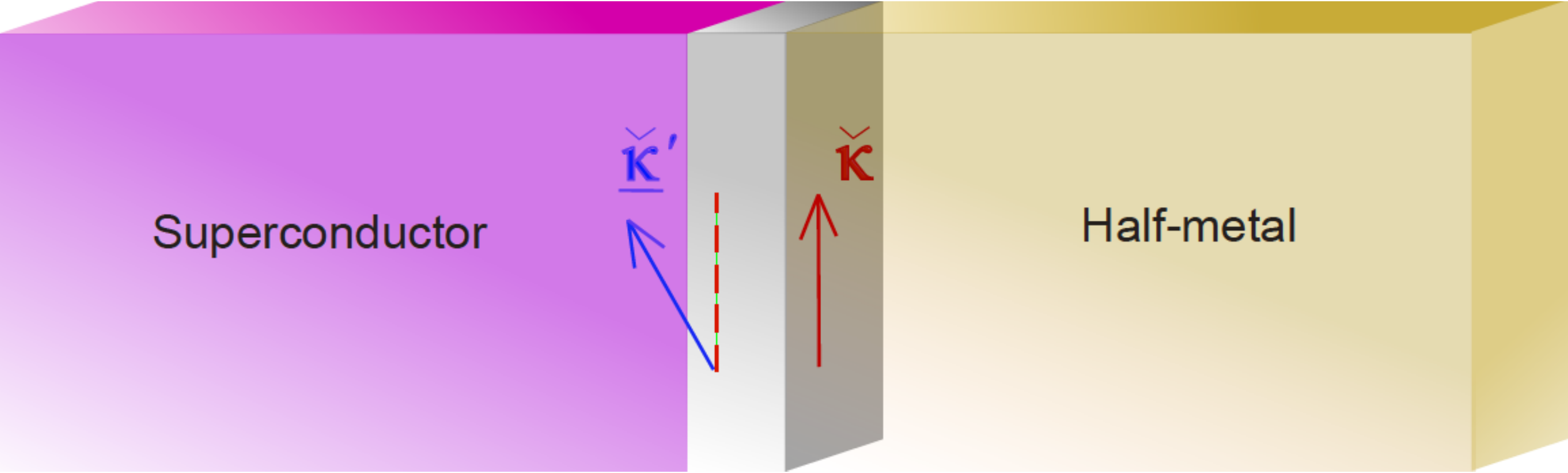}
}
\caption{
A superconductor/half-metal bilayer with a magnetically inhomogeneous barrier region. The magnetization direction associated with the spin-dependent phase-shifts occurring on the superconducting side (described by the matrix $\underline{\check \kappa}'$) does not in general align with the magnetization direction associated with the transmission of quasiparticles across the barrier (described by the matrix $\check \kappa$). 
}
\label{fig:model}
\end{figure}

The problem of a superconductor in proximity contact with a half-metallic ferromagnet has been studied within the frameworks of Eilenberger equations \cite{Eschrig04,Eschrig03,eschrig_nphys_08,Kopu04,Eschrig09,Galaktionov08,Lofwander10,Grein10}, Bogoliubov-de Gennes equations \cite{Halterman09, Linder10, Kupferschmidt11,Wilken12}, recursive Green function methods \cite{Asano07}, circuit theory \cite{Braude07}, within a magnon assisted tunneling model \cite{Takahashi07}, and in the quantum limit \cite{Beri09}.
Various experiments on superconductor/half-metal devices have been reported, both for layered systems involving high-temperature superconductors \cite{Sefrioui03,Pena04,Kalcheim12,Visani12} and in diffusive structures involving conventional superconductors \cite{Keizer06,Anwar10,Sprungmann10,Anwar11,Anwar12,Yates13,Singh15}.
An important consequence of the new boundary conditions in Eq.~\eqref{newBC} is that half-metals can now be incorporated in the Usadel equation, appropriate to describe the second class of experiments mentioned above, whereas there previously existed no suitable boundary conditions to do so. Consider first a superconductor/half-metal bilayer with the interface located at $x=0$ (see Fig. \ref{fig:model}).

The superconductor is assumed to have a thickness well exceeding the superconducting coherence length. Our expansion parameters are the spin-dependent reflection phase shifts at the superconducting side of the interface, $\underline{\vartheta}_{ll'}$, and the tunneling probabilies ${\cal T}_{nl}$.
For calculating triplet components in the half-metal
it is sufficient to expand the solution for the Green function in the superconductor
up to linear order, and the solution for the Green function in the half-metal up to quadratic order.  The zeroth order term in the superconductor is pure spin-singlet, and the first order term pure spin-triplet. Thus, up to inlcuding first order we can assume a bulk singlet order parameter, not affected by the interface scattering (corrections to the singlet order parameter arise only in second order in $\underline{\vartheta}_{ll'}$ and ${\cal T}_{nl}$). 
For future reference, we define the quantities $c\equiv\cosh(\nu)=-\i\frac{E}{\Omega }$, $s\equiv\sinh(\nu)=\i\frac{|\Delta|}{\Omega }$ with $\nu= \mbox{atanh}(|\Delta|/E)$, $\Omega=\sqrt{|\Delta|^2-E^2}$, and denote the SC phase as $\theta$.
We find for the triplet component $\underline{F}_{t0}$ in the superconductor
\begin{equation}
\underline{F}_{t0}(x)= \i 
\frac{\underline{\cal G}^\phi cs}{\sigma_{\rm SC}{\cal A}q} 
e^{\i\theta}
e^{-q|x|}(\underline{\vec{m}}'\cdot \vec{\sigma}) \i\sigma_y
\end{equation}
with 
the normal-state conductivity $\sigma_{\rm SC}=2e^2N_{\rm SC}D_{\rm SC}$ in the superconductor
($N_{\rm SC}$ and $D_{\rm SC}$ are the normal-state density of states per spin projection at the Fermi level and the diffusion constant, respectively), contact area ${\cal A}$,
and $q=\sqrt{2\Omega/\hbar D_{\rm SC}}$.

In the half-metal (width $d$), only spin-$\uparrow$ particles have a non-zero density of states at the Fermi level. 
In the spirit of quasiclassical theory of superconductivity, a strong exchange field
is incorporated not in the transport equation, but directly in the band structure which is integrated out at the quasiclassical level \cite{grein_prl_09,Grein10}, leaving only parameters like diffusion constant, and normal state density of states at the Fermi level for each itinerant spin band. For transport in a half-metallic ferromagnet, this means one must just include one spin-band with diffusion constant $D_{\rm HM}$ in the Usadel equation.
Thus, only the elements $G_{\uparrow\uparrow}$ and $F_{\uparrow\uparrow}$ exist in the Green function $\check{{\bf G}}$ of the half-metal. 
As we expand in the tunneling probability, we can (for energies well exceeding the Thouless energy $\hbar D_{\rm HM}/d^2$ of the half-metal) use the linearized Usadel equation,
\begin{equation}
\hbar D_{\rm HM} \partial_x^2 F_{\uparrow\uparrow}+ 2\i E F_{\uparrow\uparrow} = 0.
\end{equation}
Since there is only one anomalous Green function in the half-metal, we omit the spin indices for brevity of notation and define $F\equiv F_{\uparrow\uparrow}$. The general solution is $F(x)=Ae^{\i k x}+Be^{-\i kx}$ with $A,B$ being complex coefficients to be determined from the boundary conditions, and $k=\sqrt{2\i E/\hbar D_{\rm HM}}$. At the vacuum edge of the half-metal $(x=d)$, we have $\partial_x F=0$. 
At the interface between the superconductor and half-metal, the boundary conditions for $F$ from the half-metallic side is obtained from Eqs. \eqref{newBC}-\eqref{Gchi2}
with ${\cal P}_{nl} = 1$. 
Note that for ${\cal P}_{nl} = 1$, we have ${\cal G}_{\underline\chi}^0={\cal G}_{\underline\chi}^1={\cal G}_{\underline\chi}^{\rm MR} \equiv {\cal G}_{\underline\chi}$
as well as ${\cal G}^0={\cal G}^1={\cal G}^{\rm MR} $. We find that in order to obtain a non-vanishing proximity effect, it is necessary that the magnetization direction associated with transmission across the barrier ($\check{\kappa}$) and spin-dependent phase-shifts picked up on the superconducting side of the interface $(\underline{\check \kappa}')$ are different. We set $\check{\kappa} = \check{\sigma}_z$ since the barrier magnetization determining the transmission properties is expected to be dominated by the half-metal magnetization which points in the $z$-direction. The boundary condition for $F$ at $x=0$ reads:
\begin{equation}
\sigma_{\rm HM}{\cal A}\partial_x F = 2\i 
\; {\cal G}_{\underline{\vartheta}}\;
cs e^{\i\theta}(\underline{m}'_x-\i \underline{m}'_y), \quad
{\cal G}_{\underline{\vartheta}}= 2\mathcal{G}_{\underline\chi}
+ \frac{\underline{{\cal G}}^\phi {\cal G}^0 }{\sigma_{\rm SC}{\cal A}q}
\end{equation}
with the normal-state conductivity $\sigma_{\rm HM}=e^2N_{\rm HM}D_{\rm HM}$ in the half-metal 
($N_{\rm HM}$ is the normal-state density of states at the Fermi level),
and
the conductance ${\cal G}_{\underline{\vartheta}}$ contains two terms: 
$2\mathcal{G}_{\underline\chi}$ which is proportional to
$\sum_{nl} \underline{\vartheta}_{ll} {\cal T}_{nl}$,
and a second term containing $\underline{{\cal G}}^\phi {\cal G}^0$ which is proportional to $(\sum_l \underline\vartheta_{ll}) (\sum_{nl'} {\cal T}_{nl'})$.
Moreover, $\underline{m}'_x$ and $\underline{m}'_y$ are the normalized components of a possible 
misaligned barrier moment compared to the magnetization of the half-metal. We have taken
 this into account by writing:
\begin{equation}
\underline{\hat \kappa}' = \underline{m}'_x \left(\begin{array}{cc}
\sigma_x & 0 \\
0 & \sigma_ x\\
\end{array}\right)_{\! \rm ph} + \underline{m}'_y \left(\begin{array}{cc}
\sigma_y & 0 \\
0 & \sigma_y^\ast\\
\end{array}\right)_{\! \rm ph}  + \underline{m}'_z \left(\begin{array}{cc}
\sigma_z & 0 \\
0 & \sigma_ z\\
\end{array}\right)_{\! \rm ph}
\end{equation}
Inserting the general solution of $F$ into the boundary conditions, one arrives at the final result for the proximity-induced superconducting correlations $F$ in the half-metal:
\begin{equation}
F(x) = -\frac{2\cosh[\i k(x-d)]}{\sinh(\i kd)} 
\frac{{\cal G}_{\underline{\vartheta}} cs}{\sigma_{\rm HM}{\cal A}k} e^{\i\theta}(\underline{m}'_x-\i \underline{m}'_y).
\end{equation}
This is the first time the Usadel equation has been used to describe the proximity effect in a superconductor/half-metal structure. Several observations can be made from the above expression. 
For small $E$ the energy factors $c \propto E$ in the numerator and $k^2\propto E$
in the denominator cancel, such that the proximity-effect, if present, happens even at $E=0$.
The proximity-effect is seen to be non-zero only if spin-dependent scattering phases at the superconducting side of the interface are present, and at the same time their quantization axis $\underline{\kappa}'$ is misaligned with that of the transmission amplitudes, $\kappa $.
The reason for this is that phase-shifts on the half-metallic side are irrelevant on the quasiclassical level, because they are spin-scalar
(only spin-$\uparrow$  particles have a finite density of states there). 
On the other hand, the phase-shifts $\underline\vartheta_{nn}$
on the superconducting side 
have two consequences: they 
are responsible for an $\vec{S}\cdot \vec{\underline{m}}=0$ spin-triplet component on that side of the interface (where $\vec{S}$ is the spin vector of the Cooper pair), and they
affect also transmission amplitudes. As a consequence, during transmission the quantization axis $\underline\kappa'$
can be rotated into the $S_z=\pm 1$ spin triplet components which are allowed to exist in the half-metal \mbox{if} spin-flip processes exist at the interface (\eg due to some misaligned interface moments). 
This is exactly the reason for why $F$ also depends on $\underline{m}'_x$ and $\underline{m}'_y$ whereas it is independent on the barrier moment $\underline{m}'_z$: only a barrier moment with a component perpendicular to the magnetization of the half-metal can create spin-flip processes which rotate the $\vec{S}\cdot\vec{\underline{m}}=0$ into the $S_z=\pm 1$ components, and thus $F$ also vanishes if $\underline{m}'_x=\underline{m}'_y=0$.

Another important observation that can be made from the above expression is that a misaligned barrier moment effectively renormalizes the superconducting phase. Using spherical coordinates, we may write $\underline{m}'_x-\i \underline{m}'_y = \sin\underline{\Theta}' e^{-\i\underline{\varphi }'}$ where $\underline{\varphi}'$ is the azimuthal angle describing the orientation of the barrier moment in the $xy$-plane. Thus, the effective phase becomes $\theta \to \theta-\underline{\varphi }'$. To see what consequence this has in terms of measurable quantities, we proceed to consider a Josephson junction with a half-metal by replacing the vacuum boundary condition at $x=d$ with another superconductor. Solving for the anomalous Green function $F$ in the same way as above, we may compute the supercurrent flowing through the system via the formula
[see Eq. \eqref{diffcurrentstrong}]:
\begin{equation}
I = \frac{eN_{\rm HM}D_{\rm HM}\mathcal{A}}{8} \int^{\infty}_{-\infty} \mbox{d}E \mbox{Tr}\{ \hat \tau_3 (\check{\bf G}_{\rm HM}\partial_x\check{\bf G}_{\rm HM})^K\}.
\end{equation}
Here, 
Tr denotes a trace over 2$\times $2 Nambu-Gor'kov space. 
After some calculations, one arrives at the result:
\begin{equation}
\label{I0}
I = I_0
\sin\underline\Theta'_L\sin\underline\Theta'_R \sin(\theta_R-\theta_L + \underline{\varphi}'_L - \underline{\varphi}'_R),
\end{equation}
where $I_0$ is a lengthy expression depending on parameters such as the width $d$ of the half-metal and the temperature $T$ (and which vanishes unless ${\cal G}_{\underline{\vartheta}}^L$ and ${\cal G}_{\underline{\vartheta}}^R$ are non-zero).
To be general, we have allowed the spin-dependent phase-shifts for each superconductor and the barrier moment at each interface to be different, indicated by the notation '$L$' and '$R$' for left and right. 
We find that $I_0$ is negative, giving rise to a $\pi$-Josephson junction behavior for the case of $\underline{\varphi}'_L = \underline{\varphi}'_R$.
Expression \eqref{I0}
is consistent with the ballistic case result of Refs.  \cite{eschrig_nphys_08,Eschrig09,Eschrig07} and shows
 how a finite supercurrent will appear in a ring geometry even in the absence of any superconducting phase
difference, $\theta_R-\theta_L=0$, if the barrier moments are misaligned in the plane 
perpendicular to the junction, $\underline{\varphi}'_L-\underline{\varphi}'_R\neq 0$.
A similar effect was also reported via circuit theory for a diffusive system \cite{Braude07}, however not due to spin-dependent scattering phase shifts but due to some ``leakage terms''. 
Within our formalism, we thus obtain a so-called $\phi_0$ Josephson junction behavior 
\cite{Josephson62,Geshkenbein86,Krive04,Buzdin08,Reynoso08}
with $\phi_0=(\pi+\underline{\varphi}'_L - \underline{\varphi}'_R)$mod$(2\pi)$.

The above framework can be readily generalized to cover strongly spin-polarized ferromagnets
 building on the same idea as Ref.  \cite{grein_prl_09}. For a sufficiently large spin-splitting, the $\uparrow$- and $\downarrow$-conduction bands can be treated separately
 in the bulk with a separate Usadel equation for $F_{\uparrow\uparrow}$ and 
$F_{\downarrow\downarrow}$. These would then only couple via interface scattering and the strong exchange field would only enter by having different normal-state density of states $N_\uparrow $, $N_\downarrow $ and diffusion coefficients $D_\uparrow $, $D_\downarrow $ of the spin-bands in each separate Usadel equation.

\section{Conclusions}

We have derived new sets of boundary conditions for Usadel theory of superconductivity, appropriate for spin-polarized interfaces. We present a general solution of the problem appropriate for arbitrary transmission, spin-polarization, and spin-dependent scattering phases. 
The explicit equations for the most general set of boundary conditions are given in Eqs. \eqref{def1}-\eqref{god}, \eqref{def01}-\eqref{go}, and \eqref{comm}-\eqref{gl_naz}. With the solution of this long-standing problem we anticipate a multitude of practical implementations in future to tackle superconducting systems that involve strongly spin-polarized materials. We have applied the general set of equations to various special cases important for practical use.
We derived boundary conditions for an interface between a superconductor and a ferromagnetic insulator valid for arbitrary spin dependent scattering phases, Eq. \eqref{FI}. This extends previous work of Ref. \cite{cottet}, which was restricted to small scattering phases. Using an exact series expansion of the general set of boundary conditions, Eqs. \eqref{Lie}-\eqref{seriesgi}, we have obtained a perturbation series for the boundary conditions appropriate for such an interface, which allows for channel off-diagonal scattering and channel-dependent spin quantization axes, Eqs. \eqref{I3}-\eqref{three}.
For the tunneling limit, we have presented a new set of boundary conditions appropriate for arbitrary spin polarization, non-trivial spin texture across the interface, and allowing for channel off-diagonal scattering, Eqs. \eqref{mainBCa1}-\eqref{mainBCa2}. Neither of these three ranges of validity has been covered previously. As an application we then proceed to give a theoretical foundation of the boundary conditions used in Refs. \cite{Machon1,Machon2,Bergeret12}, Eqs. \eqref{newBC}-\eqref{Gfi}, which we have generalized for channel off-diagonal scattering and non-trivial spin texture across the interface. One central result of the application of our formalism is the extension of these relations to second order, including the important mixing terms between transmission and spin-dependent scattering phases. These terms, Eqs. \eqref{newBC2}-\eqref{Gchi2} generalize the corresponding terms from Ref. \cite{cottet} to arbitrary spin polarization, possible nontrivial spin-texture across the interface, and channel off-diagonal scattering.
We have demonstrated the application of the new set of boundary conditions by treating a diffusive superconductor/half-metal proximity junction and a diffusive superconductor/half-metal/superconductor Josephson junction. In the latter case we found a realization of a $\phi_0$-junction. We are confident that our boundary conditions will advance the field of superconducting spintronics considerably.

\section*{Acknowledgments}
ME acknowledges financial support from the Lars Onsager committee during his stay as Lars Onsager Professor at NTNU,
as well as support from the UK EPSRC under grant reference EP/J010618/1.
ME also benefited from fruitful discussions at the Aspen Center of Physics and within the Hubbard Theory Consortium. He thanks in particular Mikael Fogelstr\"om for valuable discussions.
AC acknowledges financial support from the ANR-NanoQuartet [ANR12BS1000701] (France).
WB acknowledges useful discussions with Peter Machon and financial support from the DFG through 
BE 3803/03 and SPP 1538, and from the Baden-W\"urttemberg-Foundation through the Network of Competence ``Functional Nanostructures''. 
JL was supported by the ``Outstanding Academic Fellows'' programme at NTNU and Norwegian Research Council grants no. 205591 and no. 216700, and acknowledges support from the Onsager committee at NTNU and by the COST Action MP-1201 ``Novel Functionalities through Optimized Confinement of Condensate and Fields''.

\appendix
\section{Singular Value Decomposition of Scattering Matrix}
\label{app1}

We perform a singular value decomposition of the reflection and transmission matrices (with dimensions $n\times n$ for $\hat S_{11}$, $m\times m$ for $\hat S_{22}$, $n\times m$ for $\hat S_{12}$, and $m\times n$ for $\hat S_{21}$)
\begin{equation}
\label{PD}
\hat {\bf S}=
\left(
\begin{array}{cc}
\hat S_{11} & \quad \hat S_{12} \\  \hat S_{21} & -\hat S_{22}
\end{array}
\right)_{\! \lr}
=
\left(
\begin{array}{cc}
URV^\dagger & W T\breve Z^\dagger \\ \breve W\breve TZ^\dagger &
-\breve U \breve R \breve V^\dagger
\end{array}
\right)_{\! \lr}.
\end{equation}
Here $U,V,W,Z,\breve U,\breve V, \breve W, \breve Z$ are unitary matrices,
and the $R,T,\breve R,\breve T$ contain the real and non-negative 
singular values in the main diagonal and are zero otherwise. I.e.,
$T^\dagger=T^T$ and $\breve T^\dagger=\breve T^T$,
$R^\dagger=R$ and $\breve R^\dagger=\breve R$. We assume that the singular values are sorted from smallest to largest in $R$ and $\breve R$, and from largest to smallest in $T$ and $\breve T$.
We introduce the unitary matrices
$\Phi= W^\dagger U$, $\Psi = Z^\dagger V$ 
$\breve \Phi =\breve W^\dagger \breve U$, and $\breve \Psi =\breve Z^\dagger \breve V$.
In terms of those, unitarity of the matrix $\hat {\bf S}$ requires that
(we denote for simplicity the unit matrices $1_{n\times n}$ and $1_{m\times m}$ with the same symbol 1; the dimension is clear from the context)
\begin{eqnarray}
(1-R^2) &=& \Phi^\dagger T T^\dagger \Phi =
\Psi^\dagger \breve T^\dagger \breve T  \Psi \\
(1-\breve R^2)&=& \breve \Phi^\dagger  \breve T \breve T^\dagger  \breve \Phi =
\breve \Psi^\dagger  T^\dagger T \breve \Psi .
\end{eqnarray}
We see that $1-R^2$ and $1-\breve R^2$ contain the eigenvalues of the Hermitian matrices on the right hand sides of the equations, which requires that these eigenvalues coincide with the values in the diagonal matrices $TT^\dagger$, $\breve T^\dagger \breve T$, $\breve T\breve T^\dagger$, and $T^\dagger T$, respectively. 
Thus, with the sorting arrangement done above,
the relations $(1-R^2)=TT^\dagger = \breve T^\dagger \breve T$ and
$(1-\breve R^2)=\breve T\breve T^\dagger=T^\dagger T$ hold.
Because all singular values of $T$ are real, this means that
$\breve T= T^\dagger $,
$R=\sqrt{1-T T^\dagger }=\sqrt{1-\breve T^\dagger \breve T }$,
$\breve R=\sqrt{1-T^\dagger T}=\sqrt{1-\breve T\breve T^\dagger }$,
and $R\breve T^\dagger= T\breve R $, $R T=\breve T^\dagger\breve R$.
Furthermore, the unitary matrices $\Phi$ and $\Psi $ commute with $R$ and the unitary matrices $\breve \Phi $ and $\breve \Psi $ commute with $\breve R$.
In particular, those matrices are block diagonal, with block sizes given by the degeneracy of the singular values in $R$ and $\breve R$, respectively.
The remaining unitarity requirements, using the above findings, reduce to
\begin{eqnarray}
\Phi \Psi^\dagger (T \breve R) &=& (T \breve R) \breve \Psi \breve \Phi^\dagger \\
\Psi \Phi^\dagger (RT) &= & (RT) \breve \Phi \breve \Psi^\dagger .
\end{eqnarray}
That means that for the blocks corresponding to non-zero reflection singular values
the above two equations lead to the one condition $\Phi^\dagger T \breve \Psi = \Psi^\dagger T \breve \Phi $. 
If there are no zero-reflection singular values then,
remembering that $\Phi$ commutes with $R$ and $\breve \Psi $ with $\breve R$,
\begin{eqnarray}
\label{SVD0}
\hat {\bf S}&=&
\left(
\begin{array}{cc} U \Phi^\dagger & 0\\ 0& \breve U \breve \Psi^\dagger \end{array}
\right)_{\! \lr}
\left( \begin{array}{cc}
R& \quad T \\ T^\dagger &
-\breve R
\end{array} \right)_{\! \lr}
\left(
\begin{array}{cc} \Phi V^\dagger & 0\\ 0& \breve \Psi \breve V^\dagger \end{array}
\right)_{\!\lr}. \quad
\end{eqnarray}
The blocks with zero-reflection singular values can be treated separately, and 
it is easily seen that the singular value decomposition of the scattering matrix has the general form
\begin{eqnarray}
\label{SVD}
\hat {\bf S}&=&
\left(
\begin{array}{cc} {\cal U} & 0\\ 0& \breve{\cal U} \end{array}
\right)_{\!\lr}
\left( \begin{array}{cc}
\sqrt{1-T T^\dagger }& T \\ T^\dagger &
-\sqrt{1-T^\dagger T}
\end{array} \right)_{\!\lr}
\left(
\begin{array}{cc} {\cal V}^\dagger & 0\\ 0& \breve{\cal V}^\dagger \end{array}
\right)_{\! \lr} 
\end{eqnarray}
with unitary matrices ${\cal U}$, $\breve{\cal U}$, ${\cal V}$, and $\breve{\cal V}$. The decomposition is not unique.

\section{Polar Decomposition of Scattering Matrix}
\label{app2}
An important feature of the above representation is that the center matrix
is Hermitian. If we only require this property of the central part, but
not necessarily diagonality of the $m\times n$ matrix $T$, then
we can find an entire class of transformations that keep this property.
We define ${\cal R}D \breve{\cal R}^\dagger =T $ with unitary matrices ${\cal R}$ and $\breve{\cal R}$.  Then
\begin{eqnarray}
\hat {\bf S}&= &
\left(
\begin{array}{cc} {\cal UR} & 0\\ 0& \breve{\cal U}\breve{\cal R} \end{array}
\right)_{\!\!\lr}\!\!
\left( \begin{array}{cc}
\sqrt{1-D D^\dagger }& D\\ D^\dagger &
-\sqrt{1-D^\dagger D}
\end{array} \right)_{\!\!\lr} \!\!
\left(
\begin{array}{cc} {\cal R}^\dagger {\cal V}^\dagger & 0\\ 0& \breve{\cal R}^\dagger \breve{\cal V}^\dagger \end{array}
\right)_{\!\!\lr}
\nonumber\\
\end{eqnarray}
where $D$ is now an $n\times m$ matrix that is not necessarily diagonal anymore.
Consider now some special cases.
First, we chose ${\cal R}={\cal V}^\dagger$, $\breve{\cal R}=\breve{\cal  V}^\dagger $. Then
\begin{equation}
\hat {\bf S}=
\left(
\begin{array}{cc} {\cal UV}^\dagger  & 0\\ 0& \breve{\cal U} \breve{\cal V}^\dagger \end{array}
\right)_{\!\lr}
\left( \begin{array}{cc}
\sqrt{1-C' {C'}^\dagger }& C'\\ {C'}^\dagger &
-\sqrt{1-{C'}^\dagger C'}
\end{array} \right)_{\!\lr}
\end{equation}
with $C'={\cal V} T \breve{\cal V}^\dagger $ gives a polar decomposition of the reflection parts of the scattering matrix $\hat {\bf S}$. Similarly, ${\cal R}={\cal U}^\dagger$, $\breve{\cal R}=\breve{\cal  U}^\dagger $ leads to
\begin{equation}
\hat {\bf S}=
\left( \begin{array}{cc}
\sqrt{1-C C^\dagger }& C\\ C^\dagger &
-\sqrt{1-C^\dagger C}
\end{array} \right)_{\!\lr}
\left(
\begin{array}{cc} {\cal UV}^\dagger  & 0\\ 0& \breve{\cal U} \breve{\cal V}^\dagger \end{array}
\right)_{\!\lr }
\end{equation}
with $C={\cal U} T \breve{\cal U}^\dagger ={\cal UV}^\dagger C' (\breve{\cal U} \breve{\cal V}^\dagger )^\dagger $. 
We can also chose a decomposition in the form
\begin{eqnarray}
\hat {\bf S}&=&
\left(
\begin{array}{cc} {\cal UV}^\dagger  & 0\\ 0& 1 \end{array}
\right)_{\!\!\lr }\!\!
\left( \begin{array}{cc}
\sqrt{1-C'' {C''}^\dagger }& C''\\ {C''}^\dagger &
-\sqrt{1-{C''}^\dagger C''}
\end{array} \right)_{\!\!\lr }\!\!
\left(
\begin{array}{cc} 1  & 0\\ 0& \breve{\cal U} \breve{\cal V}^\dagger \end{array}
\right)_{\!\!\lr }
\nonumber \\
\end{eqnarray}
with $C''={\cal V} T \breve{\cal U}^\dagger $, or other decompositions. 

These decompositions are unique when there are no zero-reflection singular values.
This means, that under the conditions of no zero-reflection channels ${\cal UV}^\dagger $ and $\breve{\cal U}\breve{\cal V}^\dagger $ are
uniquely defined, as the matrices $C$ and $D$ are. 
The unique unitary matrices ${\cal UV}^\dagger $ and $\breve{\cal U} \breve{\cal V}^\dagger $ are
the surface scattering matrices ${\cal S}$ and $\breve{\cal S}$, 
Eq. \eqref{SV0}.
\section{Parameterization of scattering matrix}
\label{app3}

We now turn to a useful parameterization of the transmission matrix $C$. We note that with the definition
\begin{equation}
C=\left(1+tt^\dagger \right)^{-1} 2t
\end{equation}
we obtain
\begin{eqnarray}
\left( \begin{array}{cc}
\sqrt{1-C C^\dagger }& C\\ C^\dagger &
-\sqrt{1-C^\dagger C}
\end{array} \right)_{\!\lr } =
\left( \begin{array}{cc}
\hat r & \quad \hat d \\ \hat d^{\dagger} & -\breve{\hat r}
\end{array} \right)_{\!\lr }
\end{eqnarray}
with
\begin{eqnarray}
\hat r &=&
\left(1+tt^\dagger \right)^{-1}
\left(1-tt^\dagger \right) \\
\breve{\hat r} &=&
\left(1+t^\dagger t\right)^{-1}
\left(1-t^\dagger t \right) \\
\hat d &=& \left(1+tt^\dagger \right)^{-1} 2t .
\end{eqnarray}
To connect with the main text, see equations \eqref{C}-\eqref{tau}.
Furthermore, if $t=u\theta v^\dagger $ is a singular decomposition 
for $t$, then $C=u[(1+\theta^2)^{-1}2\theta]v^\dagger $ is a singular decomposition of $C$. Conversely, if $C=u \delta v^\dagger $ is a singular decomposition for $C$, then $t=u [(1-\sqrt{1-\delta^2})/\delta ]v^\dagger $ is a singular decomposition for $t$. 
If $0<\theta <1 $ then $0<\delta <1$ and vice versa.
Thus, the parameterization in terms of $t$ is equivalent to that in terms of $C$. 

\section{Expansion to third order of expression \eqref{expansion}}
\label{thirdorder}
To third order we obtain from Eq. \eqref{expansion}
\begin{eqnarray}
\check{\cal I}^{(1)} &=& \i\left[ \check K , \check G\right]\\
\check{\cal I}^{(2)} &=& 
-\frac{1}{2}\left[ \check K \stackrel{2}{,} \check G\right] +
\i\left[ \check K , (\check G +\i\pi \check 1)\otimes (\check G')^{(1)} \otimes \check G  \right] \qquad \\
\check{\cal I}^{(3)} &=& 
-\frac{\i}{6}\left[ \check K \stackrel{3}{,} \check G\right] 
-\frac{1}{2}\left[ \check K \stackrel{2}{,} (\check G +\i\pi \check 1)\otimes (\check G')^{(1)} \otimes \check G  \right] \nonumber \\
&&+\i\left[ \check K , (\check G +\i\pi \check 1)\otimes (\check G')^{(2)} \otimes \check G  \right] 
\nonumber \\
&&+\i\left[ \check K ,(\check G +\i\pi \check 1)\otimes (\check G')^{(1)} \otimes \check G \otimes (\check G')^{(1)} \otimes \check G \right] 
\end{eqnarray}
and
\begin{eqnarray}
\label{G1}
(\check G')^{(1)} =  -\frac{\i}{2\pi^2} \left[ \check K , \check G\right], \quad
(\check G')^{(2)} =  -\frac{1}{4\pi^2} \left[ \check K \stackrel{2}{,} \check G\right] .\quad
\end{eqnarray}
This can be simplified further noting 
\begin{eqnarray}
\check G\otimes (\check G')^{(1)} &=&-(\check G')^{(1)}\otimes \check G, \\
2\pi^2 (\check G')^{(1)}\otimes (\check G')^{(1)} &=& -\left\{ (\check G')^{(2)},\check G\right\}, \\
(\check G +\i\pi \check 1)\otimes \check G&=& \quad \i\pi (\check G +\i\pi \check 1), \\
2\pi^2 \left[\check K\stackrel{n}{,} (\check G')^{(1)} \right] &=& -\i \left[\check K\stackrel{n+1}{,} \check G\right], \\
4\pi^2 \left[\check K, (\check G')^{(2)} \right] &=& -\left[\check K\stackrel{3}{,} \check G \right], 
\end{eqnarray}
yielding Eq. \eqref{I3} of the main text.

\section{Expansion of matrix current for finite transmission}
\label{secondorder}

From section \ref{series} we obtain the following expressions to second order in the spin dependent reflection phases and in the transmission probability:
\begin{eqnarray}
&&\check{\cal I}^{(1)}=-2\pi \i 
\left[ \check t^o,\check g_0^o \right]^{(1)}_\otimes
+\i \left[ \check K ,\check G \right] , \\
&&\check{\cal I}^{(2)}=-2\pi \i 
\left[ \check t^o,\check g_0^o\right]^{(2)}_\otimes
+\i \left[ \check K ,(\check g^i)^{(1)}\right] 
-\frac{1}{2} \left[ \check K \stackrel{2}{,} \check G\right] ,\qquad
\end{eqnarray}
with
\begin{eqnarray}
\left[ \check t^o,\check g_0^o \right]^{(1)}_\otimes&=& \left[ \check \tau_0 \underline{\check G} \check \tau_0^\dagger ,\check G \right]_\otimes \\
\left[ \check t^o,\check g_0^o \right]^{(1)}_\otimes&=& \left[ \check G_1 ,\check G \right]_\otimes \\
\left[ \check t^o,\check g_0^o \right]^{(2)}_\otimes&=& \left[ \check \tau_0 (\underline{\check g}_0^i)^{(1)} \check \tau_0^\dagger ,\check G \right]_\otimes 
+
\left[ \check G_1,(\check g_0^o)^{(1)} \right]_\otimes  \nonumber \\
&&+ \check G_1 \otimes \check G \otimes
\left[ \check G_1 ,\check G \right]_\otimes + 
\left[ \check G_1 ,\check G \right]_\otimes
\otimes \check G \otimes \check G_1
\nonumber \\
&&+\frac{\i}{2} \left( \left[ \left[ \check K,\check G_1 \right],\check G\right]_\otimes + \left[ \tau_0 \left[\underline{\check K},\underline{\check G}\right] \tau_0^\dagger,\check G \right]_\otimes \right)
,
\end{eqnarray}
and 
\begin{eqnarray}
(\check g^i)^{(1)}&=&(\check G+\i\pi\check 1) \otimes \left( \left[ \check G_1 ,\check G\right]_\otimes
+ (\check G')^{(1)}\otimes \check G \right) \qquad\\
(\underline{\check g}_0^i)^{(1)}&=&(\underline{\check G}+\i\pi\underline{\check 1}) \otimes (\underline{\check G}')^{(1)}\otimes \underline{\check G }\\
(\check g_0^o)^{(1)}&=&(\check G-\i\pi\check 1) \otimes (\check G'')^{(1)}\otimes \check G 
\end{eqnarray}
with $(\check G'')^{(1)}=-(\check G')^{(1)}$ from Eq. \eqref{G1}. 
Collecting everything together, we obtain the result shown in Eqs. \eqref{mainBC1}-\eqref{mainBC2} of the main text.

\section{Term of second order in transmission probability}
\label{I4}
For completeness we present here the expression of order ${\cal T}_{nl}^2$:
\begin{eqnarray}
I_4&=& 
 {\cal G}_{4}^0\; \underline{\check{\bf G}} \otimes \check{\bf G} \otimes  \underline{\check{\bf G}} + {\cal G}_{4}^1\;
\check \kappa \underline{\check{\bf G}} \check \kappa \otimes \check{\bf G} \otimes \check \kappa \underline{\check{\bf G}} \check \kappa \nonumber \\
&+&{\cal G}_{4}^{\rm MR} \left(
\underline{\check{\bf G}} \otimes \check{\bf G} \otimes \check \kappa \underline{\check{\bf G}} \check \kappa + \check \kappa \underline{\check{\bf G}} \check \kappa \otimes \check{\bf G} \otimes \underline{\check{\bf G}} \right)+ 
{\cal G}_{4}^{\rm MR'}
\left\{\check \kappa ,\underline{\check{\bf G}} \right\} \otimes \check{\bf G} \otimes \left\{\check \kappa ,\underline{\check{\bf G}} \right\} \nonumber \\
&+&{\cal G}_{4}^{\rm mix} \left(\underline{\check{\bf G}} \otimes \check{\bf G} \otimes \left\{\check \kappa ,\underline{\check{\bf G}} \right\} + \left\{\check \kappa ,\underline{\check{\bf G}} \right\} \otimes \check{\bf G} \otimes  \underline{\check{\bf G}}\right)
\nonumber\\
&+& {\cal G}_{4}^{\rm mix'} \left(
\check \kappa \underline{\check{\bf G}} \check \kappa \otimes \check{\bf G}  \otimes \left\{\check \kappa ,\underline{\check{\bf G}} \right\} + \left\{\check \kappa ,\underline{\check{\bf G}} \right\} \otimes \check{\bf G} \otimes \check \kappa \underline{\check{\bf G}} \check \kappa \right)
\end{eqnarray}
with
\begin{eqnarray}
{\cal G}_{4}^0 &=& \frac{1}{8} {\cal G}_q \sum_{nln'l'}
p_{nln'l'}
{\cal T}_{nl} {\cal T}_{n'l'} 
\Big( 1 + \sqrt{1-{\cal P}_{nl}^2} \;\Big)
\Big( 1 + \sqrt{1-{\cal P}_{n'l'}^2} \; \Big) \\
{\cal G}_{4}^1 &=& \frac{1}{8} {\cal G}_q \sum_{nln'l'}
p_{nln'l'}
{\cal T}_{nl} {\cal T}_{n'l'} 
\Big( 1 - \sqrt{1-{\cal P}_{nl}^2} \;\Big)
\Big( 1 - \sqrt{1-{\cal P}_{n'l'}^2} \; \Big) \\
{\cal G}_{4}^{\rm MR} &=& \frac{1}{8} {\cal G}_q \sum_{nln'l'}
p_{nln'l'}
{\cal T}_{nl} {\cal T}_{n'l'}
\Big( 1 + \sqrt{1-{\cal P}_{nl}^2} \;\Big)
\Big( 1 - \sqrt{1-{\cal P}_{n'l'}^2} \;\Big) \\
{\cal G}_{4}^{\rm MR'} &=& \frac{1}{8} {\cal G}_q \sum_{nln'l'}
p_{nln'l'}
{\cal T}_{nl} {\cal T}_{n'l'} {\cal P}_{nl}{\cal P}_{n'l'}
\\
{\cal G}_{4}^{\rm mix} &=& \frac{1}{8} {\cal G}_q \sum_{nln'l'}
p_{nln'l'}
{\cal T}_{nl}{\cal T}_{n'l'}{\cal P}_{n'l'}
\Big( 1 + \sqrt{1-{\cal P}_{nl}^2} \; \Big)\\
{\cal G}_{4}^{\rm mix'} &=& \frac{1}{8} {\cal G}_q \sum_{nln'l'}
p_{nln'l'}
{\cal T}_{nl}{\cal T}_{n'l'}{\cal P}_{n'l'}
\Big( 1 - \sqrt{1-{\cal P}_{nl}^2} \;\Big)
\end{eqnarray}
with $p_{nln'l'}\equiv \delta_{nn'}+\delta_{ll'}-\delta_{nn'}\delta_{ll'}$,
arising from averaging over the typical phase factor $e^{\i(\psi_{nl}-\psi_{n'l}+\psi_{n'l'}-\psi_{nl'})}$ of spin-scalar transmission phases. The channel-diagonal case follows from setting $n=l=n'=l'$ and $p_{nnnn}=1$.

\end{document}